\documentclass[]{mn2e}
\usepackage{graphicx}
\usepackage{lscape}
\usepackage{deluxetable}
\usepackage{epsfig}
\pdfoutput=1

\newcommand{\HI}{\mbox{H\,{\sc i}}}
\newcommand{\MgII}{\mbox{Mg\,{\sc ii}}}
\newcommand{\MgI}{\mbox{Mg\,{\sc i}}}
\newcommand{\FeII}{\mbox{Fe\,{\sc ii}}}
\newcommand{\CIV}{\mbox{C\,{\sc iv}}}
\newcommand{\CII}{\mbox{C\,{\sc ii}}}

\newcommand{\MnII}{\mbox{Mn\,{\sc ii}}}
\newcommand{\OI}{\mbox{O\,{\sc i}}}
\newcommand{\AlII}{\mbox{Al\,{\sc ii}}}
\newcommand{\SiIV}{\mbox{Si\,{\sc iv}}}
\newcommand{\SiII}{\mbox{Si\,{\sc ii}}}
\newcommand{\NV}{\mbox{N\,{\sc v}}}
\newcommand{\CIIs}{\mbox{C\,{\sc ii$^*$}}}

\title[Quasar Sightlines Through M31]{Probing the Extended Gaseous 
Regions of M31 with Quasar Absorption Lines\thanks{Based on observations made
with the NOAO 2.1-m and the NASA/ESA {\it Hubble Space Telescope}.}}
\author[S. M. Rao et al.]
{\parbox[t]{\textwidth}{\raggedright 
Sandhya M. Rao$^{1}$\thanks{E-mail: srao@pitt.edu},
Gendith Sardane$^{1}$,
David A. Turnshek$^{1}$,
David Thilker$^{2}$,
Rene Walterbos$^{3}$,
Daniel Vanden Berk$^{4}$\thanks{Visiting Astronomer,
Kitt Peak National Observatory, National Optical Astronomy
Observatory, which is operated by the Association of Universities for
Research in Astronomy, Inc. (AURA), under cooperative agreement with
the National Science Foundation.},
and Donald G. York$^{5}$}
\vspace*{6pt}\\
$^{1}$Department of Physics and
Astronomy and PITTsburgh Particle physics, Astrophysics, and Cosmology Center 
(PITT PACC),\\
 University of Pittsburgh, Pittsburgh, PA 15260\\
$^{2}$Center for
Astrophysical Sciences, Johns Hopkins Univ., 3400 North Charles
St., Baltimore, MD 21218\\
$^{3}$Department of Astronomy, New Mexico State University,
MSC 4500, Box 30001, Las Cruces, NM 88003\\
$^{4}$Physics Department, St. Vincent College, Latrobe, PA 15650\\
$^{5}$Department of
Astronomy and Astrophysics, University of Chicago, Chicago, IL 60637
}

\begin{document}

\date{}

\pagerange{\pageref{firstpage}--\pageref{lastpage}} \pubyear{2013}

\maketitle

\label{firstpage}

\begin{abstract}

We present  {\it Hubble Space Telescope - Cosmic Origins Spectrograph}
spectra of ten quasars located behind 
M31, selected to investigate the properties of gas associated with its
extended disk and high velocity clouds (HVCs). The sightlines have
impact parameters ranging between $b= 13$ kpc and $112$ kpc. No
absorption is detected in the four sightlines selected to sample  any
extended disk (or halo) gas that might be present in the outer regions
of  M31 beyond an impact parameter of  $b>57$ kpc. Of the six
remaining sightlines, all of which lie at $b<32$ kpc and within the
$N_{HI}= 2\times 10^{18}$ cm$^{-2}$ boundary of the \HI\ disk of M31,
we detect low-ionization absorption at M31 velocities along four of
them (three of which include \MgII\ absorption).  We also detect
\MgII\ absorption from an HVC. This HVC sightline does  not pass
through the 21 cm disk of M31, but we detect additional \MgII\
absorption at velocities distinct from the HVC that presumably  arises
in the halo.  We find that along sightlines where both are detected,
the velocity location of the low-ion gas tracks the peak in 21 cm
emission.  High-ionization absorption is detected along the three 
inner sightlines, but not along the three outer sightlines, for which
\CIV\ data exist.

As inferred from high-resolution 21 cm
emission line maps  of M31's disk and extended regions, only one of
the sightlines may be capable of harboring a damped Ly$\alpha$ system,
i.e., with $N_{HI} \ge 2\times10^{20}$ cm$^{-2}$.  This sightline
has impact parameter  $b= 17.5$ kpc, and we detect both low- and
high-ion absorption lines associated with it.  

The impact parameters
of our observed sightlines through M31 are similar to the impact
parameters of galaxies identified with \MgII\ absorbers at redshifts
$0.1<z<1.0$ in a 2011 study by Rao {\it et al.}  However, even if we
only count cases  where absorption due to M31 is detected, the \MgII\
$\lambda$2796 rest equivalent width  values are significantly smaller.
In comparison, moderate-to-strong \MgII\ absorption from Milky Way gas
is detected along all  ten sightlines. Thus, this study indicates that
M31 does not present itself as an absorbing galaxy which is $typical$
of higher-redshift galaxies inferred to give rise to moderate-strength
quasar absorption lines. M31 also appears not to possess an extensive
large gaseous cross section, at least not along the direction of its
major axis. 

\end{abstract}


\section{Introduction}
\label{intro}

The standard paradigm for metal-line absorption systems in quasar
spectra is that they arise in the extended gaseous halos/disks of
galaxies well beyond their observable optical radii. However, with the
exceptions afforded by gravitationally-lensed quasars, rarely is there
more than one sightline passing in the vicinity of a galaxy.  As such,
the study of quasar absorption lines arising in extended gas
associated with the great spiral galaxy in Andromeda (M31) represents
a unique opportunity. M31's large extent on the sky  means that many
quasar sightlines should intercept its extended gas.  For example, the
$N_{HI}=1.9\times 10^{18}$ atoms cm$^{-2}$ 21 cm emission contour around
M31, as derived from the data discussed by Thilker et al. (2004), is
approximately $5.0 \times 1.5$ square degrees on the sky  (see Figure
1). We list some properties of M31 in Table 1.  Quasar surveys have
shown that there are as many as 18 quasars per square degree brighter
than $g\sim 20$ at $z\la 2.6$ (Richards et al. 2009, Abraham et
al. 2012). Thus, there are likely to be on the order of 135 such
quasars behind M31 within the boundaries of its observed 21 cm
emission, and a factor of several more in its extended gaseous disk
and halo regions. However, until now, quasar absorption lines have
never been  successfully used to study the extended gas of M31 because
of the lack of sufficiently-bright, identified quasars.

Two of the most recognizable signatures of metal lines in quasar
spectra are the \MgII\ $\lambda\lambda$2796,2803 and \CIV\
$\lambda\lambda$1548,1550 doublets, which have been studied in
numerous quasar absorption-line surveys. The first comprehensive
study which demonstrated that galaxies at large impact parameters
exist along the sightlines to low-redshift \MgII\ absorbers was by
Bergeron and collaborators, e.g., Bergeron \& Boiss\'e (1991).  They estimated that
the average \MgII\ radius of a spherical gaseous envelope surrounding
an $L^*$ galaxy is $R^* \sim$ 3.5 to 5.0$R_H$ ($\sim$ 55 to 80 kpc) at
$z\sim0.3$ for rest equivalent widths  $W_0^{\lambda2796} \ge 0.3$
\AA, where $R_H$ is the Holmberg radius.  Others had made similar
estimates (e.g., Lanzetta et al. 1987, Lanzetta \& Bowen 1990, Steidel
1993).  The recent survey of galaxies associated with  \MgII\
absorbers at $0.1<z<1.0$ by Rao et al. (2011) showed that the gaseous extent of
\MgII-selected absorbing galaxies could be as large as 100 kpc.
At $z < 0.5$, Chen et al. (2010) find that the mean covering 
fraction for \MgII\ absorbers with $W_0^{\lambda2796} \ge 0.3$ \AA\
within $\sim$130  kpc of a 2$L^*$ galaxy (for $h=0.7$) 
is $\sim$70\%.  Therefore, if cross sections have
remained constant since $z\sim 0.5$,  then we might expect that gas giving rise to \MgII\
is likely to be present in the extended gaseous regions of M31 out to
a radius of $\sim 100$ kpc or more, assuming it is a typical absorbing
galaxy.

\begin{table}
\caption{M31 properties\label{tbl.1}}
\begin{tabular}{lcc}
\hline
\hline
Property & Value & Reference\tablenotemark{a} \\ 
\hline
RA (2000) & 00$^{\rm h}$42$^{\rm m}$44$^{\rm s}$ & 1 \\
Dec (2000) & +41$^\circ$16$\arcmin$08$\arcsec$ & 1 \\
Distance & $752 \pm 27$ kpc & 2 \\
Inclination & 78$^\circ$ & 3 \\
$v_{sys}$ &  $-306$ km/s & 3 \\
$R_{opt}$\tablenotemark{b} & 22.3 kpc & 4 \\
$m_B$ & 4.16 & 4 \\
$L_B$\tablenotemark{c} & $2.0 L_B^*$ & 4,5 \\
$R_{21cm}$\tablenotemark{d} & 33 kpc & 3 \\
$M_{virial}$ & $0.8-1.1 \times 10^{12} M_{\odot}$ & 6 \\ 
\hline
\end{tabular}
\tablenotetext{a}{References: 1. Evans et al. (2010); 2. Riess et al. (2012); 3. Corbelli et al. (2010); 
4. de Vaucouleurs et al. (1991); 5. Cool et al. (2012); 6. Tamm et al. (2012)}
\tablenotetext{b}{Optical radius at $B$-band surface brightness $\mu_B=25$ magnitudes per square arcsec.}
\tablenotetext{c}{Assuming $M_B^*=-19.92$ (Cool et al. 2012).}
\tablenotetext{d}{From the $N_{HI} = 1.9 \times 10^{18}$ atoms cm$^{-2}$ contour (Figure 1).}
\end{table}

As described in \S2, we obtained {\it Hubble Space Telescope (HST) -
Cosmic Origins Spectrograph (COS)} spectra of ten quasars
located behind M31 in order to
investigate the properties of the gas in its extended disk and high
velocity clouds (HVCs). We searched for \MgII, \CIV, and other
absorption lines to do this. In \S3 we  describe the results obtained
from each spectrum. We discuss the results in \S4 and end with a
summary and conclusions in \S5.  This study indicates that  M31 does
not present itself as an absorbing galaxy which is $typical$ of the
higher-redshift  galaxies inferred to give rise to moderate-strength
quasar absorption lines.

\section{Observations}
\label{obs}

\subsection{Existing \HI\ 21 cm Emission Observations}

Since M31 is the nearest large spiral galaxy close to the Milky Way,
it has been the subject of many observational studies. Specifically
for this work, we will make reference to several results over  the
past decade from   radio observational studies of M31's \HI\ 21 cm
emission.  These are: the Green Bank Telescope (GBT) study of Thilker
et al. (2004), which identified high-velocity clouds (HVCs) but at
lower spatial resolution than later studies; the Westerbork Synthesis
Radio Telescope (WSRT) study of Braun and Thilker (2004) which discovered
the M31-M33 \HI\ bridge, and of Westmeier et al. (2005), which focused
on obtaining  higher spatial resolution observations of HVCs; the WSRT
study of Braun et al. (2009), which  obtained observations over a wide
field at high spatial resolution; and the study of Corbelli et
al. (2010),  which smoothed the data to lower spatial resolution in
order to fit a tilted-ring model to M31's warped disk and study its
rotation curve.    At some level, all of this work was collaborative
by various members of the same group,  and in later studies they made
use of results that could be derived from earlier data sets.

\HI\ emission spectra were extracted from the Thilker et al. (2004) and
Braun et al. (2009) datacubes along the sightlines toward our target
quasars. These data, originally in units of Jy/beam, were converted to
$N_{HI}$ under the assumption of negligible \HI\ opacity.  This
conventional assumption, while recently questioned by Braun et
al. (2009) and Braun (2012) in the dense gaseous environment of the
traditional optical disk and slightly beyond, is expected to be
satisfied in the outer disk and halo environment.  A more significant
concern regarding the $N_{HI}$ from observations of emission is the
vastly different scale probed by the GBT and WSRT relative to COS. The
maximum linear spatial resolution of the high resolution 21 cm
observations is $\sim 50-100$ pc at the distance of M31.  This scale
is of order $\sim10^5$ times larger than the linear spatial scale
sampled in quasar  ``pencil-beam'' absorption-line observations, where
the pencil-beam has the scale of the UV continuum  emitting region of
the background quasar. Thus, $N_{HI}$ values derived from 21 cm
emission  observations are averaged over a much larger spatial scale
in comparison to those derived from quasar absorption-line
spectra. Nevertheless, using 21 cm observations to derive  average
$N_{HI}$ values along our sightlines, and noting the velocity range of
detected emission, provides some important information.

As an aside, we note that it would be interesting if $N_{HI}$ results
derived from M31's 21 cm emission  data could someday be compared with
$N_{HI}$ determinations from Lyman series absorption seen in the UV
spectra of background quasars.  One could then get an \HI\ column
density measurement  averaged over less than a milli-parsec region in
M31, in comparison to the $\sim$ 50 pc linear spatial  scale offered
by the radio observations. This would provide information on the
homogeneity and size scale of \HI\ absorbing regions in M31.

\subsection{Optical Discovery Spectra of Quasars behind M31}

We started this  project by developing a list of quasars in especially
desirable locations (see below) relative to M31.  These were initially
quasar candidates,  since existing catalogs generally did not include
quasars behind M31.  The quasar candidates were selected from special
plates of the SDSS, which were obtained specifically to find quasars
behind the extended regions of M31  (Adelman-McCarthy et al. 2006).
Of the 219 candidates, 108 were confirmed as quasars. Twenty-three of
the 108 were spectroscopically  confirmed during our October 2003 NOAO
2.1 m Gold Camera run at Kitt Peak. To make follow-up observations
with HST-COS (\S2.3 and \S3) more feasible, we concentrated on finding
brighter quasars. We also focused our search behind
M31's extended major axis to probe possible disk gas that could sample 
its outer rotation curve. See Figures 1 and 2, and Tables
1 and 2,  for information on their locations
relative to M31 and the  discovery spectra.  Quasars labeled 1 through
4 would sample any extended disk gas (or possibly halo gas) that is
undetected in 21 cm emission; quasars 5, 6, 8, and 9 lie  near the
edge of detected 21 cm emission; the sightline towards quasar 7 passes
through a high velocity cloud (HVC) in the circumgalactic environment
of M31; and quasar 10 lies behind the 21 cm emission \HI\ disk
as well as two other HVCs.  Importantly, owing to M31's systemic velocity
of  $-306$ km s$^{-1}$ (Corbelli et al. 2010) and its direction of
rotation, absorption originating on the southwest side of M31 will not
be confused with Galactic absorption. Consequently, quasars 1 through
4 offer the best opportunities for observing  extended disk gas and
measuring M31's rotation curve much farther out than possible with 21
cm emission observations.  Unfortunately, while obtaining information
on M31's rotation curve at large galactocentric distance was one of
the primary motivations for observing quasars 1 through 4, no M31
absorption near the expected velocity was detected in their UV spectra
(\S2.3 and \S3).  We note, however, that higher quality
observations  might yet be able to detect gas at these locations.  
Observing quasars on the extended
northeast  side of M31 was avoided because of potential confusion with
Galactic absorption.

\begin{figure*}
\includegraphics[width=6.0in]{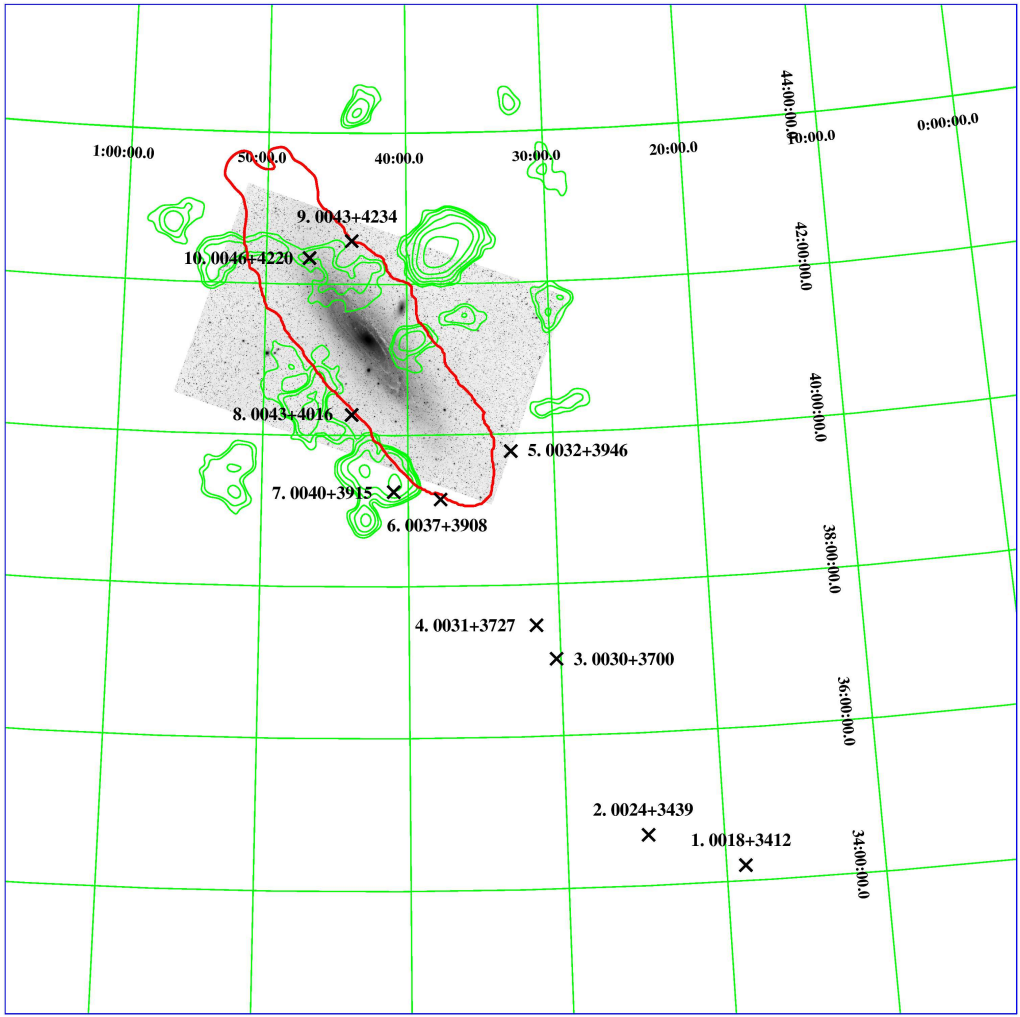}
\caption{Location of the ten quasars that were observed with {\it HST-COS}.
An optical image of M31 is shown in the background along with 21 cm
emission  maps showing the disk gas and HVCs. The red contour is 21 cm emission at
1 Jy km/s, or an \HI\ column density of $N_{HI}=1.9\times 10^{18}$ cm$^{-2}$. 
Higher column density contours interior to this are 
not shown. High velocity cloud contours from Thilker et
al. (2004) are shown in green.  The scale at the distance of M31 is 13.2
kpc deg$^{-1}$. The innermost (8. 0043+4016) and outermost (1. 0018+3412) quasars are at projected distances
(impact parameters) of $b=13.4$ kpc and $b=111.9$ kpc from M31's center. See Tables 1 and 2.}
\label{map}
\end{figure*}

\begin{figure*}
\includegraphics[width=6.0in]{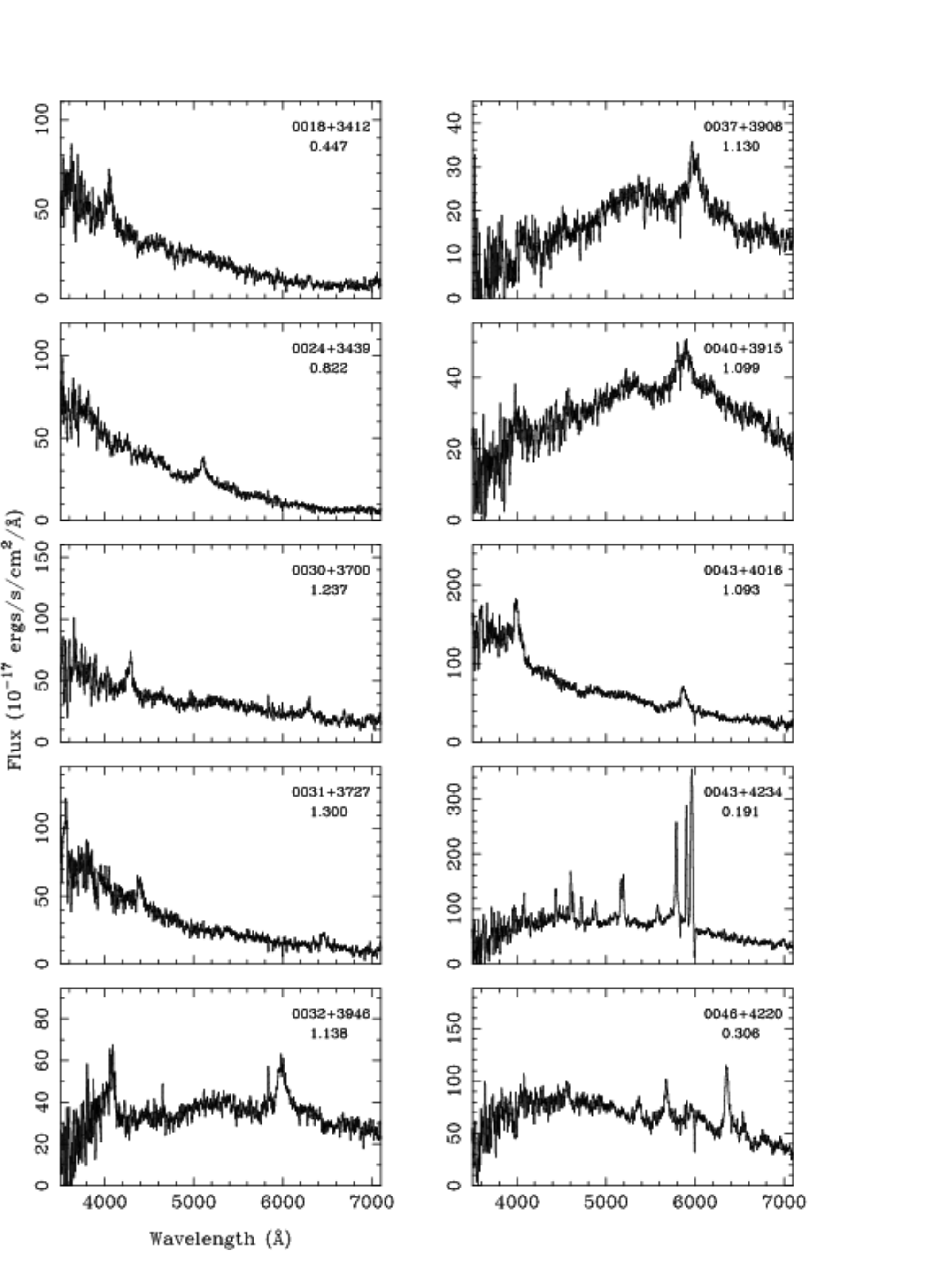}
\caption{KPNO 2.1m Gold Camera discovery spectra of the ten quasars
that were observed  with HST-COS. The quasar name and emission
redshift are noted in each panel.}
\label{KPspectra}
\end{figure*}

\begin{table*}
\caption{Quasars observed with HST-COS\label{tbl.2}}
\begin{tabular}{cccccccccl}
\hline
\hline
Quasar & RA (2000) & Dec (2000) & Map & $z_{em}$ & SDSS $u$ & b & COS G140L & COS G230L & Sightline notes\tablenotemark{c}\\
 & h m s & $^\circ$ $\arcmin$ $\arcsec$ & ID\tablenotemark{a} & & mag & (kpc)\tablenotemark{b} & Exp time (s) &
Exp time (s) & \\
\hline
0018+3412 &  00 18 47.45 & +34 12 09.6 & 1 & 0.447 & 17.7 & 111.9 & 3446    & 5646 &   extended disk \\ 
0024+3439 &  00 24 50.05 & +34 39 42.8 & 2 & 0.822 & 18.1 &  98.6 & \nodata & 5092 &   extended disk \\  
0030+3700 &  00 30 17.43 & +37 00 54.3 & 3 & 1.237 & 17.5 &  64.4 & 4054    & 3191 &   extended disk \\  
0031+3727 &  00 31 32.37 & +37 27 51.8 & 4 & 1.300 & 18.5 &  57.6 & 4455    & 2246 &   extended disk \\  
0032+3946 &  00 32 55.70 & +39 46 19.3 & 5 & 1.138 & 18.6 &  31.5 & \nodata & 10324 &   edge of 21 cm disk \\  
0037+3908 &  00 37 48.00 & +39 08 58.7 & 6 & 1.130 & 18.4 &  30.5 & \nodata & 10354 &   edge of 21 cm disk \\  
0040+3915 &  00 40 59.03 & +39 15 12.3 & 7 & 1.099 & 18.5 &  26.9 & 6293    & 11101 &   HVC + 21 cm disk edge\\ 
0043+4016 &  00 43 52.45 & +40 16 29.4 & 8 & 1.093 & 18.2 &  13.4 & 7073    & 5537 &   edge of 21 cm disk \\  
0043+4234 &  00 43 54.98 & +42 34 30.4 & 9 & 0.191 & 18.0 &  17.4 & 3555    & 6137 &   edge of 21 cm disk \\  
0046+4220 &  00 46 55.52 & +42 20 50.1 &10 & 0.306 & 18.1 &  17.5 & 2572    & 5217 &   2 HVCs + 21 cm disk \\  
\hline
\tablenotetext{a}{Quasar IDs in order of increasing RA.}
\tablenotetext{b}{Projected distance from galactic center assuming
that the center of M31 is at  (00$^{\rm h}$42$^{\rm m}$44$^{\rm s}$, +41$^\circ$16$\arcmin$08$\arcsec$) and that the
distance to M31 is 752 kpc. See Table 1.}
\tablenotetext{c}{See Figure 1.}
\end{tabular}
\end{table*}

\subsection{{\it HST-COS} UV Spectroscopy}

The {\it HST-COS} spectroscopic data were obtained during the period
July-October 2010. Table \ref{tbl.2} gives details of the quasars and
the HST-COS observations. We decided to make a broad initial
absorption-line survey in order to maximize the observed number of
metal-line transitions we could reasonably cover within our allocation
of 39 HST orbits.\footnote{Parallel imaging data were also obtained. 
These will be discussed in Thilker et al. (in prep.).} 
The aim was to reach a signal-to-noise ratio which
would enable us to detect \MgII\ and \CIV\ absorption rest equivalent
widths  commonly seen in prior, large moderate-resolution quasar
absorption-line surveys. Therefore, we did not use higher-resolution
COS gratings. However, it would indeed be worthwhile to perform
follow-up spectroscopy of a number of our detections at higher
spectral resolution and signal-to-noise ratios.
 
The COS gratings used in this study along each sightline are
specified in Table \ref{tbl.2}.  The near ultraviolet (NUV) G230L
grating has a resolution of 2 pixels or $\sim 0.82$ \AA\ at the
wavelength of the \MgII\ $\lambda\lambda$2796,2803 doublet,  which
corresponds to $\sim 87$ km s$^{-1}$ on a velocity scale.   The far
ultraviolet (FUV) G140L grating has a resolution of 7 pixels or $\sim
0.55$ \AA\ at the wavelength of the \CIV\ $\lambda\lambda$1548,1550
doublet, which corresponds to $\sim 106$ km s$^{-1}$.  Given the
redshifts of the quasars, we should note that in certain wavelength
regions  there is the possibility of contamination by Ly$\alpha$
forest absorption. For example, Ly$\alpha$ forest absorption would
potentially be visible near any  Galactic or M31 \MgII\ absorption
when the quasar's redshift is  higher than $z_{em} \sim 1.3$ (i.e., in
quasar 4) and near any Galactic or M31 \CIV\ absorption when
the quasar's redshift is higher  than $z_{em} \sim 0.27$ (i.e., in all
quasars except quasar 9).  However, according to Weymann et
al. (1998), the  incidence of Ly$\alpha$ forest absorption lines with
rest equivalent widths $\ge$ 0.24 \AA\ at these relatively low
redshifts is typically  only about one line per $30$ \AA\ (about one
line per $3200$ km s$^{-1}$), so we did not  necessarily anticipate too
much confusion due to overlapping Ly$\alpha$ forest absorption. There
might also be overlapping absorption due to unidentified metal-line
systems.  In \S3 we note instances where Ly$\alpha$ forest  absorption
or other overlapping unidentified absorption appears to be a confusing
factor.

Seven  quasars were observed with both the NUV and FUV gratings, while
three were targeted with the NUV grating alone.  These three had low
FUV fluxes based on the {\it GALaxy Evolution eXplorer (GALEX)} telescope
measurements, and so they were not
observed. The NUV grating  covers \FeII, \MnII, \MgII\ and \MgI\
transitions, while the FUV grating covers \CIV, \SiIV\ and several
lower-ion transitions, as described  in \S3.

Pipeline flux-calibrated and wavelength-calibrated spectra were used
for all the measurements, and  no additional calibrations or
re-calibrations were carried out. The wavelength scale is
heliocentric,  and measured velocity offsets relative to a transition
of interest are made on this scale.   Before making absorption-line
measurements, the FUV spectra were re-binned to two pixels per
resolution element and all spectra were normalized using an
interactive algorithm which fitted splines to a quasar's observed
continuum plus broad emission lines to derive a pseudo-continuum.  We
used the pipeline-provided standard deviation in flux to calculate the
1$\sigma$ error in the normalized flux.  When reporting errors in
equivalent width measurements, we do not include (propagate)  any
errors that might arise during the process of defining a
pseudo-continuum.

 \section{Results}

Figures $3-12$ show the pseudo-continuum-normalized spectra near the
predicted locations of metal lines along the ten sightlines, and Table
3 gives the measured metal-line absorption rest equivalent widths or
upper limits for both M31 and Galactic lines. To make these
measurements, heliocentric velocity locations  for the absorbing gas
had to be determined. The procedure  for this is discussed below and
the results on velocity offsets are given in Table 4.
 
For the low-ion transitions, the narrow \MnII\ lines (when present)
allow for  a more accurate determination of the velocity centroid of
Galactic gas since they are well-fitted by single
Gaussians. Therefore, the velocity offsets of low-ion Galactic
absorption lines are defined by  the centroids of Galactic \MnII\
$\lambda2576$ absorption for sightlines 2, 8, 9, and 10\footnote{The 
sightline 10 Galactic component is heavily blended with the M31 disk component,
as described in the discussion of sightline 10.}.  The
centroids of Galactic \MgII\ $\lambda2796$ are used to define the
velocity offsets of absorption along  other sightlines. The wavelength
interval covered by the {\it COS-FUV} spectra includes transitions due to
\SiII, \OI, \CII, \CII$^*$, \FeII, and \AlII. The centroids of these
low-ion lines were fixed at the velocities determined from either the
\MnII\ $\lambda2576$ line or \MgII\ $\lambda2796$ as indicated above.
In the panels for each
figure, dash-dot vertical lines are drawn at the determined velocity
offsets of M31 and Galactic gas.

The only high-ion transitions detected in our spectra are due to
C$^{3+}$ and Si$^{3+}$. The velocity centroids of Gaussians fitted to
the \CIV\ $\lambda1548$ lines were allowed to vary since low-ion and
high-ion absorption lines are not {\it a priori} required to have the
same  velocity centroids or line widths. The \CIV\ $\lambda1550$ line
and \SiIV\ lines were then constrained to have the same velocity
locations and widths as the \CIV\ $\lambda1548$ line, within the
uncertainties and resolution of the data. Inspection of the final fits
suggests that this was a reasonable constraint.

The 1$\sigma$ error in the normalized flux is shown in the figures as
a black dotted line.  M31 and Galactic absorption transitions that are
identified at a level of  significance $>2$$\sigma$ are indicated in
the figures by red profiles. A $>2$$\sigma$ rest equivalent
width detection threshold is an appropriate criterion for identifying
absorption  because we already know the approximate velocity location
of M31 absorption (e.g., from M31's 21 cm emission). We also searched
for significant absorption in a wider velocity window. Gaussian
profiles are fitted to detected absorption.  If more than one Gaussian
is required to fit the data, we show the individual Gaussians  as red
dashed profiles, visible above the solid red profile. In the absence
of multiple Gaussians, the red solid profile will lie on top of the
red dashed profile, and the red dashed profile will not  be
visible. However, the measurements indicated by the red dashed
profiles are what we report in Tables 3 and 4.  As noted earlier, 
the positions of most low-ion lines are fixed by the  centroid
of either the \MnII\ $\lambda2576$ or the \MgII\ $\lambda2796$ line;
however, their widths are allowed to vary in order to obtain the best
fit.  In a few cases, even the velocity offsets had to be allowed to
vary up to one  resolution element in order to obtain a  satisfactory
fit.  Also, while performing the fits, we identified some  absorption
in the spectra which were likely blends resulting from a real M31 or
Galactic absorption line plus overlapping or nearby absorption due to,
for example, Ly$\alpha$ forest absorption, some other unrelated
absorption, or even related absorption such as \CII\ $\lambda$1334.5
and \CIIs\ $\lambda$1335.7.  When this happened, we fitted Gaussians
to these nearby absorption components in order to better isolate the
M31 and Galactic absorption transition of interest. We refer to this
as deblending.  However, when we report results in Tables 3 and 4, as
noted earlier, only absorption taken to be due to the designated
transition of interest in M31 or the Galaxy is reported and shown on
the figures. Other nearby absorption lines which were fitted in order
to isolate M31 and Galactic gas are shown as green dashed Gaussian
profiles.  The identifications and measurements of M31 and Galactic
lines in the presence of confusing overlapping or nearby absorption
should be considered less secure.

When a line is not detected (i.e., the detection is $<2$$\sigma$) at
its expected velocity offset, or nearby  absorption not due to the
transition of interest appears to be present,  a red dotted Gaussian
profile with  FWHM equal to the spectrograph resolution  (i.e., $\sim
0.82$ \AA\ or $\sim 87$ km s$^{-1}$  for the NUV lines and $\sim 0.55$
\AA\ or $\sim 106$ km s$^{-1}$ for the FUV lines) is shown on the
figures to indicate the  reported upper limit.  If no overlapping or
nearby confusing absorption is present, this is just the $2\sigma$
upper limit on  equivalent width generated  from the error in
normalized flux. However, if overlapping or nearby absorption is
present, the  upper limit is determined from the strength of this
overlapping or nearby absorption. Lacking evidence that a
low-oscillator-strength transition should be present along a
particular sightline, we would attribute any significant detected
absorption as due to overlapping  absorption, and list it as an upper
limit.

In cases where the velocity of an M31 absorption line overlaps with
the velocity of a different Galactic absorption line, for example, the
M31 \CIV\ $\lambda 1550$ and the Galactic \CIV\ $\lambda 1548$ lines,
or the M31 \SiII\ $\lambda 1304$ and the Galactic \OI\ $\lambda 1302$
lines along sightlines 1, 3, and 4 (Figures 3, 5, and 6),  we
assign the absorption to the Galactic absorption system. The
measurement is listed in Table 3 only for the Galactic absorption line.

The bottom panels for sightlines 5 through 10 (Figures $7-12$) show
\HI\ 21 cm emission profiles extracted from the GBT data of Thilker et
al. (2004).  The intensities are scaled to 
accentuate the very weak emission signal from M31.  The
dash-dot horizontal line drawn in each 21 cm panel marks the location
of zero intensity. The \HI\ 21 cm emission disk of M31 extends
out to $\sim 33$ kpc as determined from the $N_{HI}=1.9\times 10^{18}$
cm$^{-2}$ column density contour (Figure 1),
and no \HI\ 21 cm  measurements exist at the positions of quasars 1
through 4. Therefore, to estimate equivalent width upper limits for
these four sightlines, we have assumed that M31's 21 cm rotation curve
is flat at large galactocentric distances and we extrapolate
the sightline 21 cm emission velocity out to the positions of quasars
1 through 4 to predict a probable velocity location of absorbing
gas. Note that M31 is nearly edge-on and inclined $\sim 78\deg$ on
the plane of the sky. Thus a very small inclination correction is
needed since $\sin(78) = 0.978$.  Then the assumption of a flat
rotation curve suggests that if metal-line absorption is present in
M31's outer regions,  we might find it near a heliocentric  velocity
location of $\sim -525$ km s$^{-1}$. This is where we determine M31
equivalent width upper limits for sightlines 1 through 4. We note that
the  choice of where to measure potential absorption in the four outer
sightlines is purely an algorithmic decision given that flat  rotation
curves exist. We also considered the Tamm et al. (2012) study which
derives a rotation curve out to a galactocentric radius of  $\sim 500$
kpc.  They employ, among other diagnostics, observations of stellar
streams (Fardal et al. 2006)  and satellite galaxies (Tollerud et
al. 2012) which yield rotational velocities of $\sim 160$ km s$^{-1}$
near the position of our outermost sightline.  This translates to a
heliocentric velocity of $-466$  km s$^{-1}$ since our outer
sightlines lie on the approaching, SW, side of M31. This is well
within one resolution element (\S 2.3) of our assumed velocity
location of $ -525$ km s$^{-1}$.  Therefore, we are confident that we
have not missed any absorption from gas in M31 along the outer four
sightlines that is above our detection limits.

Tables 3 and 4 summarize all of the measurements and upper limits,
both for M31 and the Milky Way Galaxy. A discussion of individual
sightlines follows (see Figures $3-12$), with  emphasis on what they
reveal about M31 gas. The discussions are presented in order  of
increasing sightline right ascensions. This ordering generally follows
decreasing impact  parameter, $b$, except for the last three
sightlines which all have $13 < b < 18$ kpc.  At the beginning of each
discussion  we indicate the maximum wavelength at which Ly$\alpha$
forest absorption might cause  blending and confusion,
$\lambda_{forest} \sim 1216(1+z_{em})$ \AA.

\begin{description}

\item[{\bf 1. 0018+3412 ($b=111.9$ kpc, $\lambda_{forest} < 1760$ \AA,
Fig. 3):}]  No significant M31 absorption is detected  along this
sightline, and \HI\ 21 cm emission maps of M31 do not extend  this far
out. Therefore, rest equivalent width upper limits on absorption  were
measured at $-525$ km s$^{-1}$ as described earlier. At this velocity
location, the red dotted Gaussian lines show the velocity positions
and rest equivalent widths of hypothetical unresolved absorption lines
with 2$\sigma$ levels of significance, and these are the upper limits
reported in Table 3.  Galactic absorption is clearly present.
Suspected confusion (blending) due to overlapping Ly$\alpha$ forest
absorption is apparent for the \SiII\ $\lambda$1260, \SiII\ $\lambda$1304,
\OI\ $\lambda$1302, \CII\ $\lambda$1334, and \CIV\ $\lambda$1550 Galactic absorption
lines. The method we used to measure such cases was discussed above.

\begin{figure*}
\hspace{-0.4in}
\includegraphics[width=3.2in]{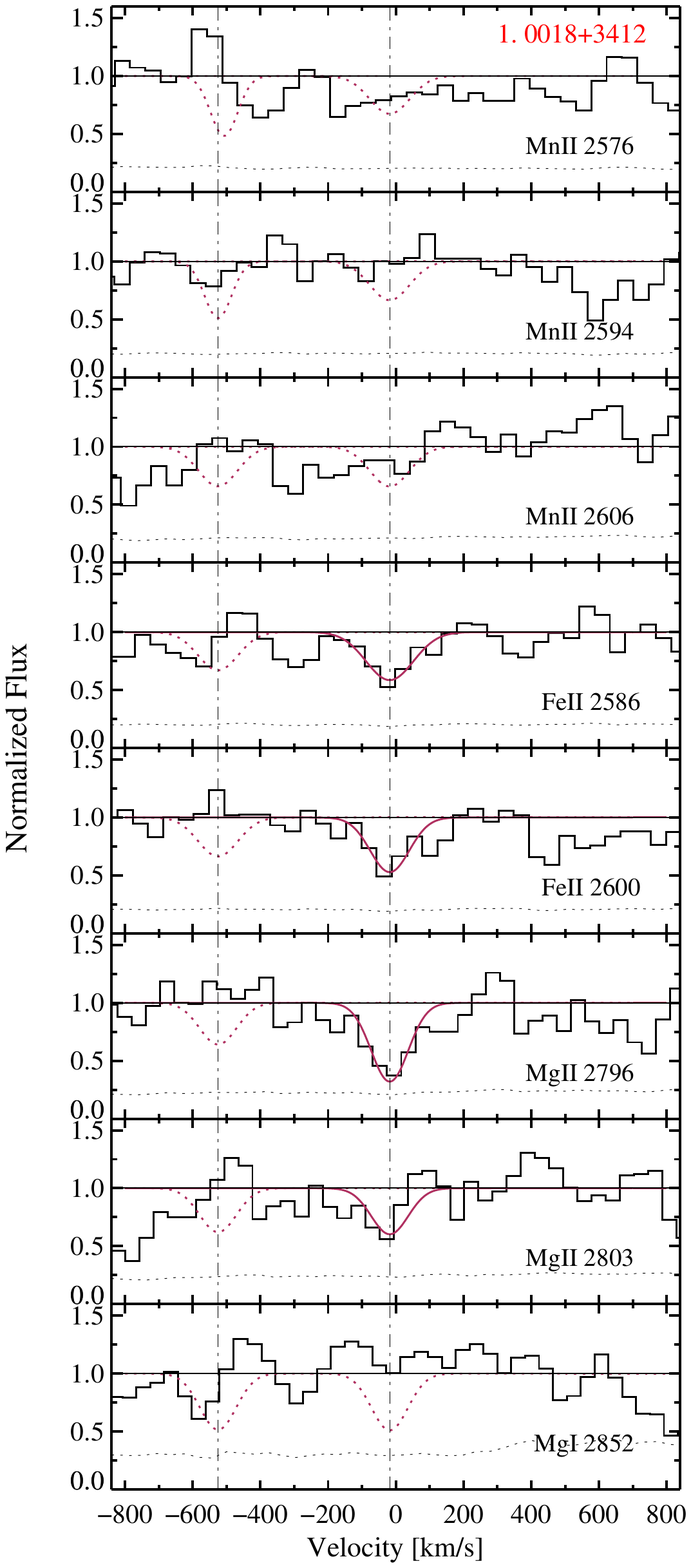}\hspace{-1.3in}
\includegraphics[width=3.2in]{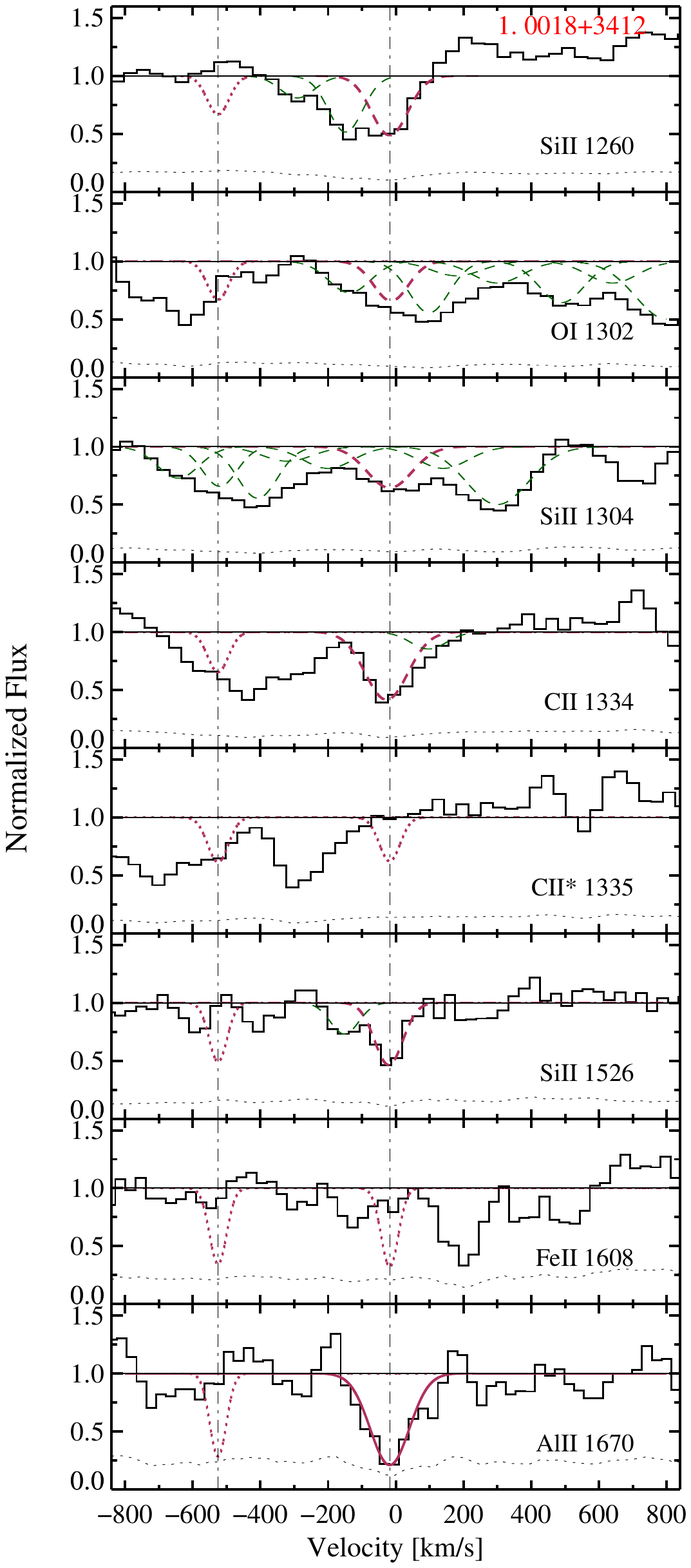}\hspace{-1.3in}
\includegraphics[width=3.2in]{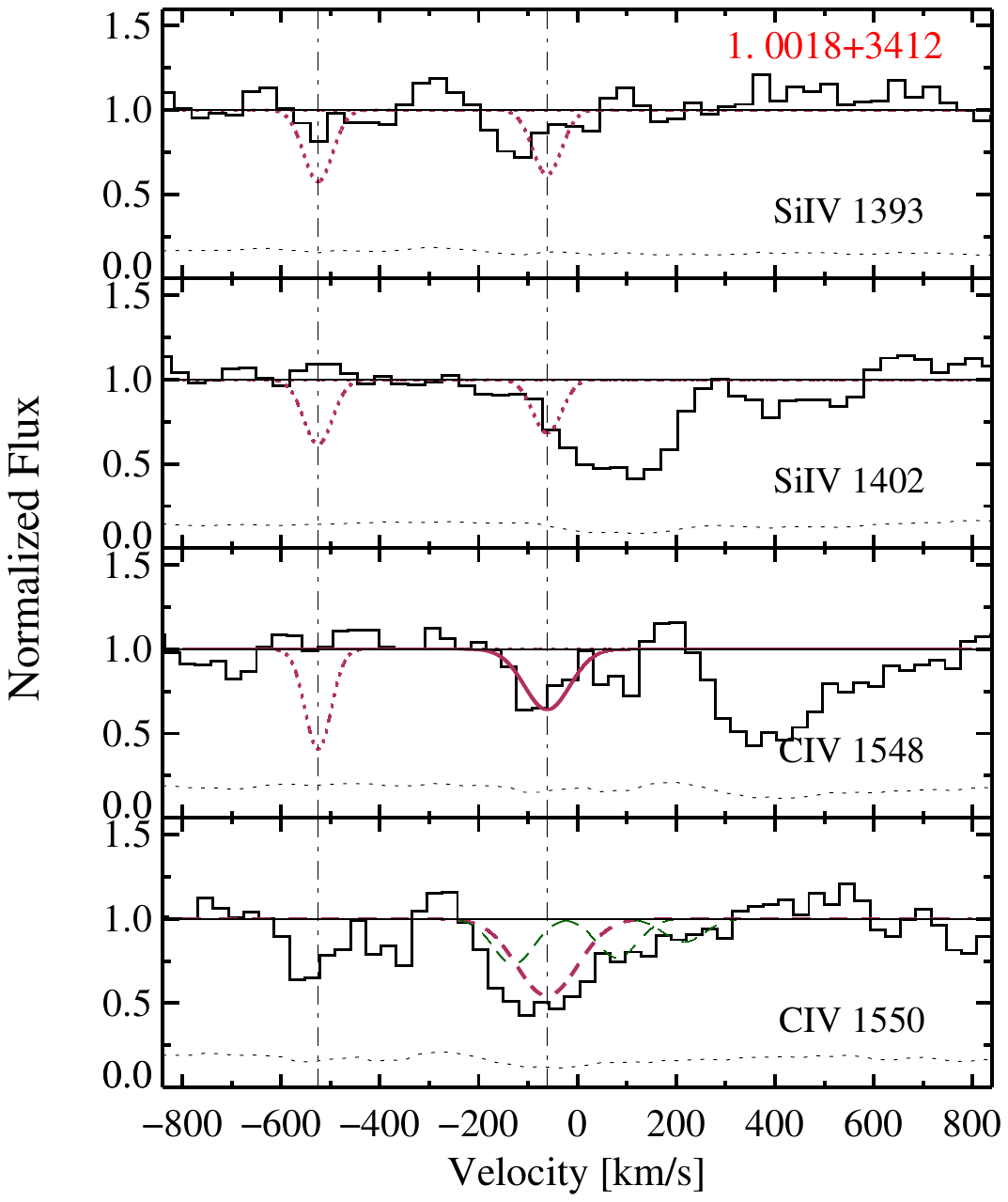}
\caption{Normalized spectra versus velocity for the labeled
transitions  in the spectra of 0018+3412. The black dotted line is the
$1\sigma$ error spectrum.  All velocities are heliocentric. The vertical
dot-dashed lines are Milky Way  (near 0 km s$^{-1}$) and M31 detected or
assumed velocities. See Table 4.  Fits to M31 and Galactic absorption lines detected at
a significance $>2\sigma$ are shown as heavy dashed (if part of a blend) or solid red lines. 
See text. Dotted red lines indicate $2\sigma$ upper limits. Green dashed lines are components
within a blend that are unrelated to M31 or Galactic absorption.
 }
\end{figure*}

\item[{\bf 2. 0024+3439 ($b=98.6$ kpc, $\lambda_{forest} < 2216$ \AA,
Fig. 4):}]  As in the previous sightline, no significant M31
absorption is detected, and \HI\ 21 cm emission maps do not extend
this far out, so   upper limits were measured  at a velocity location
of $-525$ km s$^{-1}$.  Only NUV spectra of this quasar were
obtained. Therefore, for example, the \CIV\ region was not observed. A
Galactic MnII $\lambda$2576 line is detected at a level of
significance of $\sim3\sigma$, however the two  weaker members of the
triplet are not detected at $>2\sigma$.  Galactic \MgII\ and \FeII\
absorption are clearly detected.

\begin{figure*}
\includegraphics[width=4.5in]{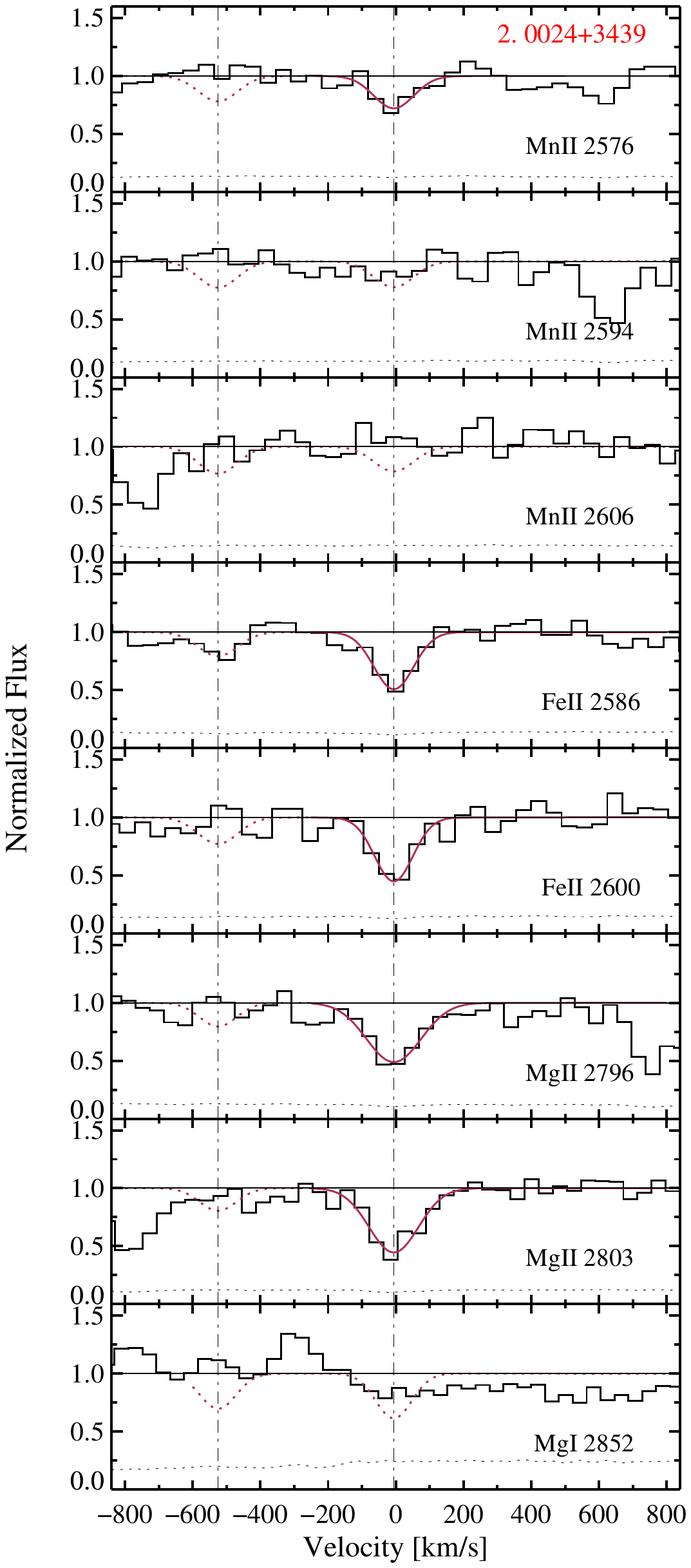}
\caption{Same as Figure 3, but for 0024+3439. No FUV spectra 
of this quasar were obtained.}
\end{figure*}

\item[{\bf 3. 0030+3700 ($b=64.4$ kpc, $\lambda_{forest} < 2720$ \AA,
Fig. 5):}]   Again, no significant absorption lines from M31 are
detected at or near $-525$ km s$^{-1}$, and the 21 cm emission maps do
not extend out this far.  Among the significant Galactic absorption
lines that are detected, the measurements of \SiII\ $\lambda$1260,
\SiIV\ $\lambda$1393, \CIV\ $\lambda$1548 and \FeII\ $\lambda$2586
were made in the presence of overlapping unrelated absorption using
the method described earlier.  While only the stronger members of the
Galactic \SiIV\ and \CIV\ doublets are  detected, the rest equivalent
width upper limits of the weaker members of these doublets  are
consistent with their expected strengths based on $f\lambda$ values.
In addition to the detected Galactic metal absorption lines,  at least
two partial Lyman limit absorption systems are present in the
spectrum.  One at $z \sim 0.5$ is clearly visible in the FUV
observation (not shown).  Based on the difference in flux level
between the FUV and NUV observations, and the presence of some  strong
Ly$\alpha$ forest absorption near and just shortward of the Ly$\alpha$
broad emission line, at least one other Lyman limit absorption system
is likely to be present at $1.21 < z < 1.24$. However, it  is not
directly visible in our observations because it falls in the
wavelength gap between  the FUV and NUV spectra.

\begin{figure*}
\includegraphics[width=3.3in]{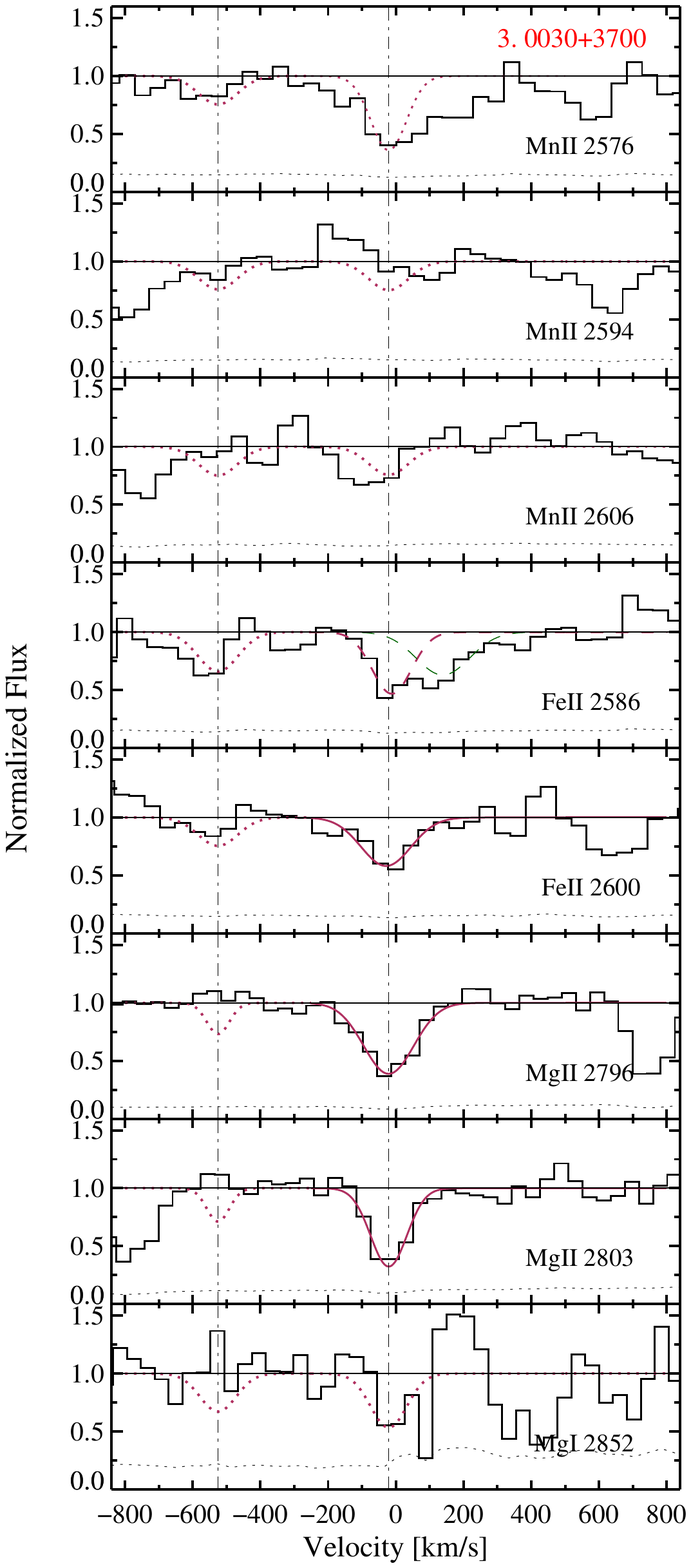}\hspace{-1.6in}
\includegraphics[width=3.3in]{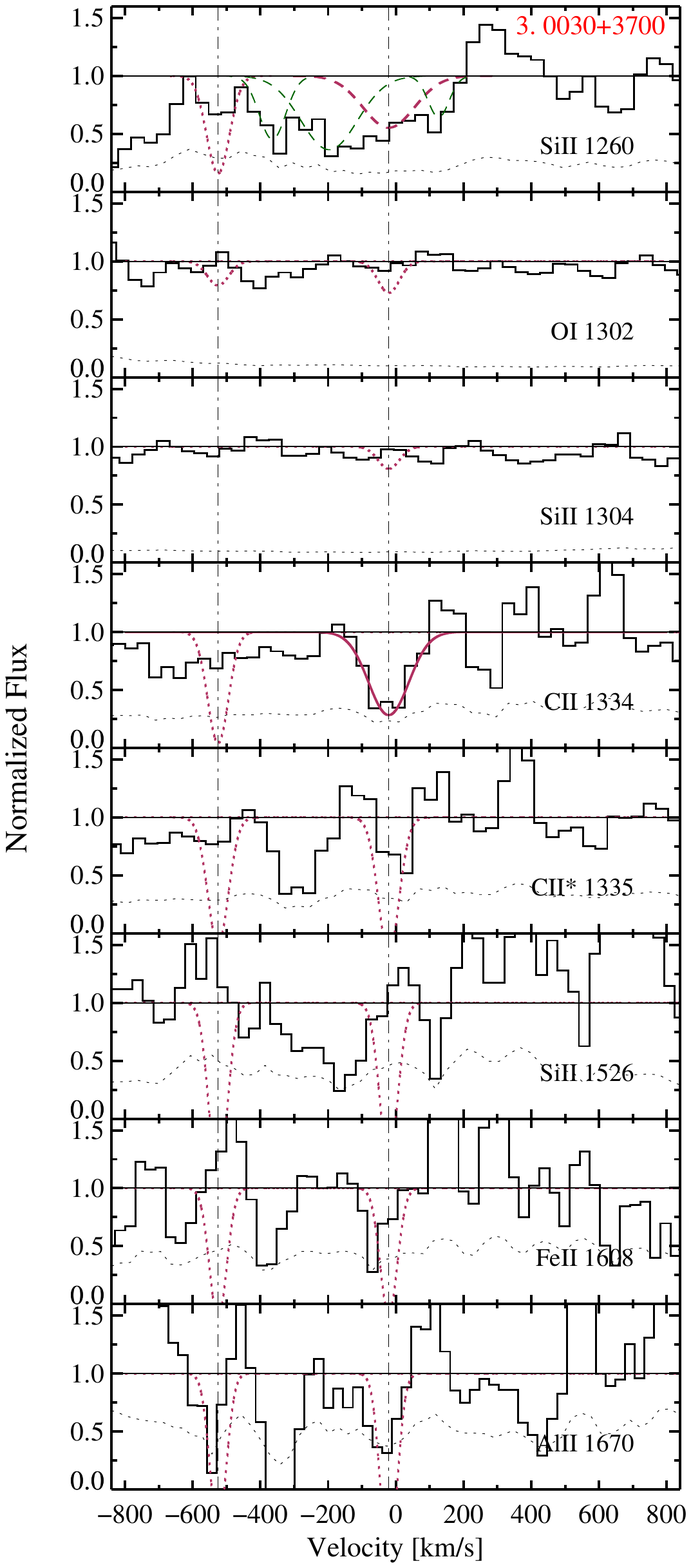}\hspace{-1.6in}
\includegraphics[width=3.3in]{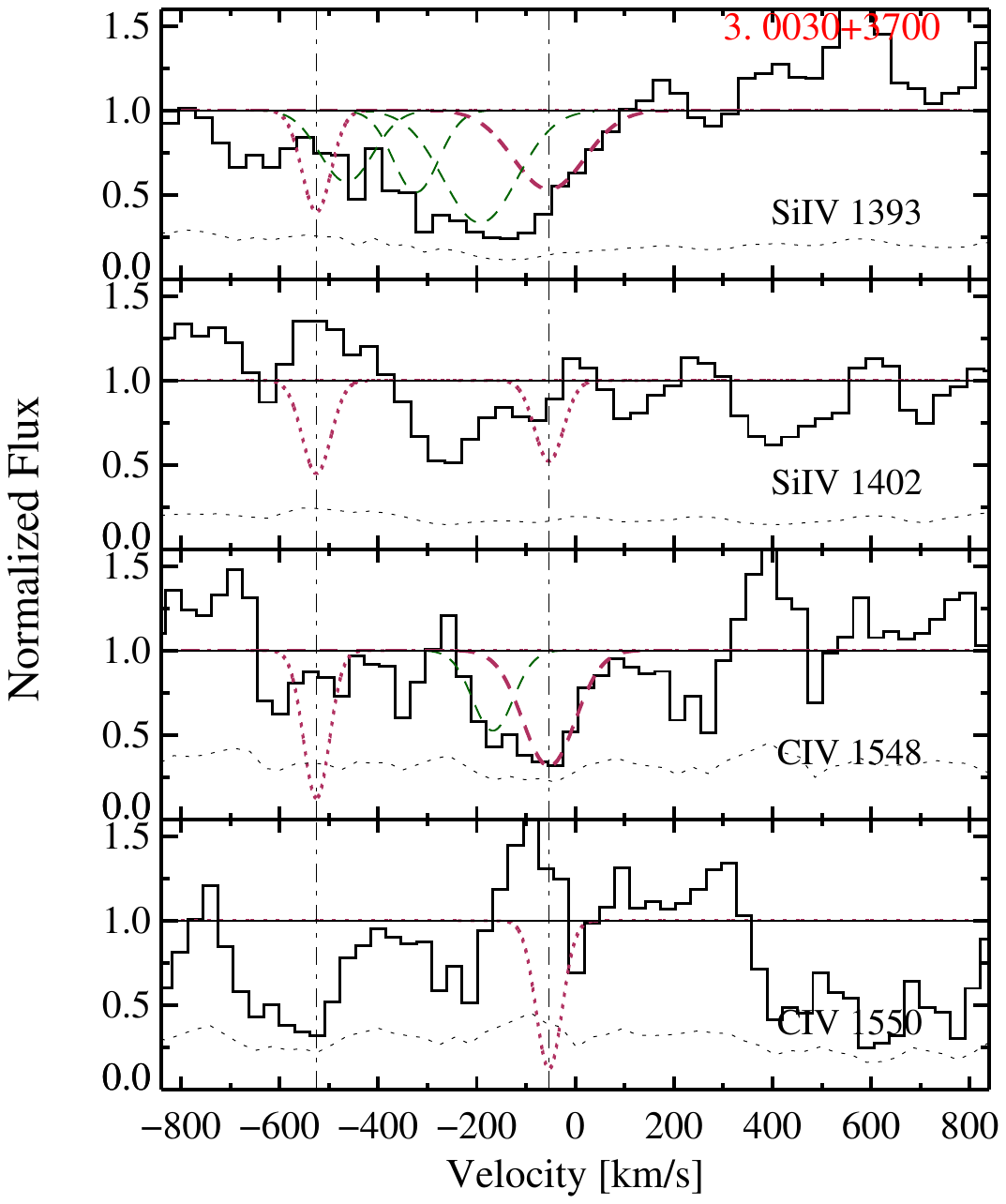}
\caption{Same as Figure 3, but for 0030+3700.}
\end{figure*}

\item[{\bf 4. 0031+3727 ($b=57.6$ kpc, $\lambda_{forest} < 2797$ \AA,
Fig. 6):}] As with the first three sightlines, no significant
absorption lines from M31 gas are seen, and the 21 cm emission map
does not extend out this far. M31 upper limits were estimated at
$-525$ km s$^{-1}$ for both the high and low ions.  Galactic
absorption is clearly detected for some transitions, but the
measurements of \SiII\ $\lambda$1260, \CII\ $\lambda$1334,  \SiII\
$\lambda$1526, \FeII\ $\lambda$1608, and \FeII\ $\lambda$2600 required 
deblending due to the presence of unrelated overlapping absorption.
 
\begin{figure*}
\includegraphics[width=3.3in]{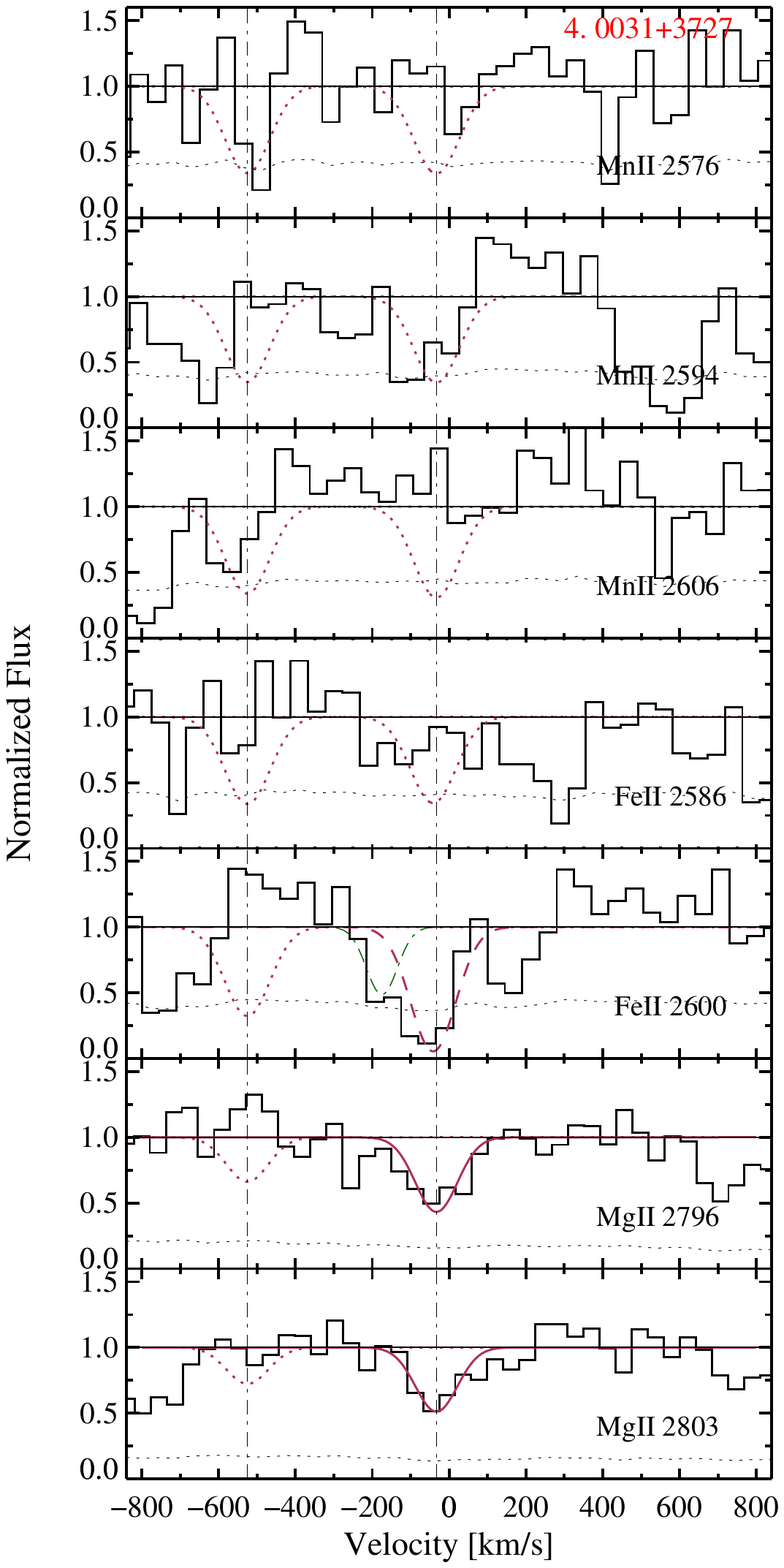}\hspace{-1.6in}
\includegraphics[width=3.3in]{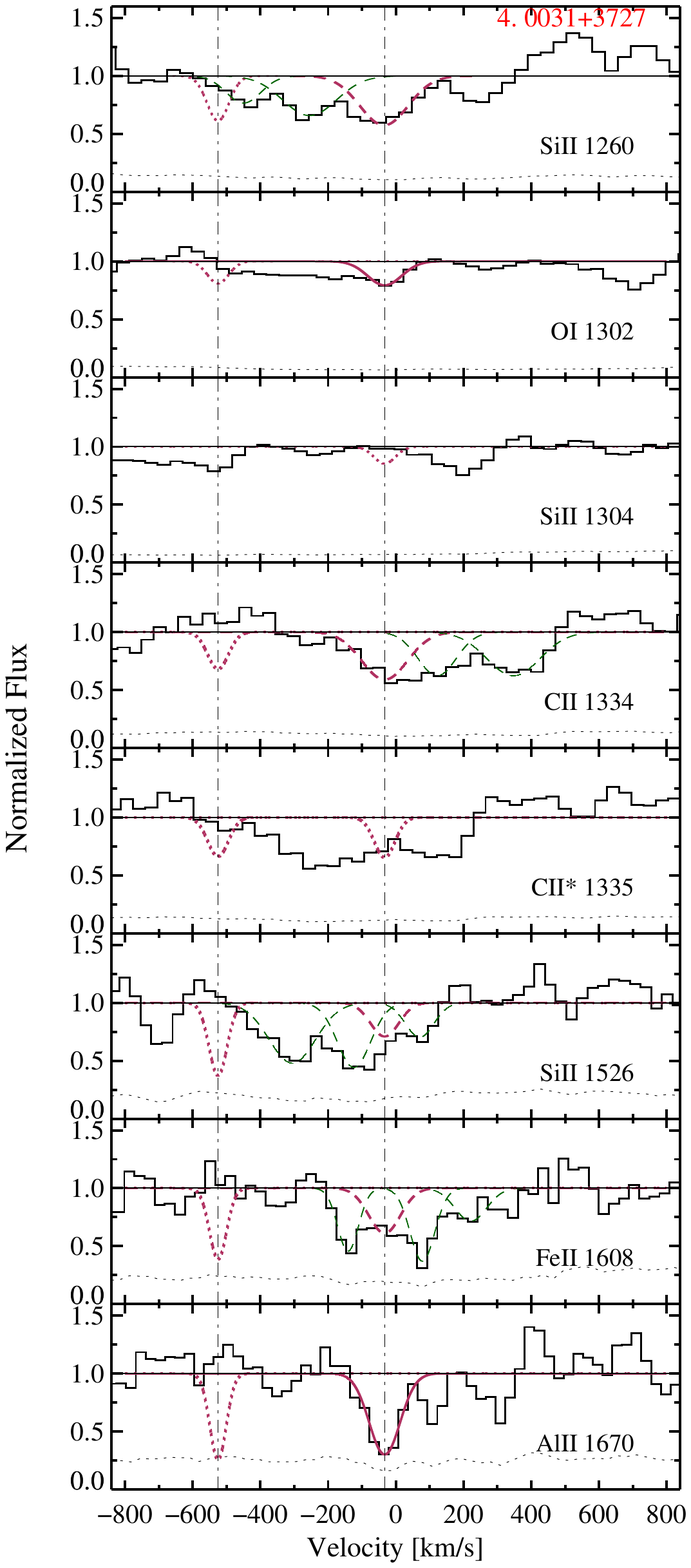}\hspace{-1.6in}
\includegraphics[width=3.3in]{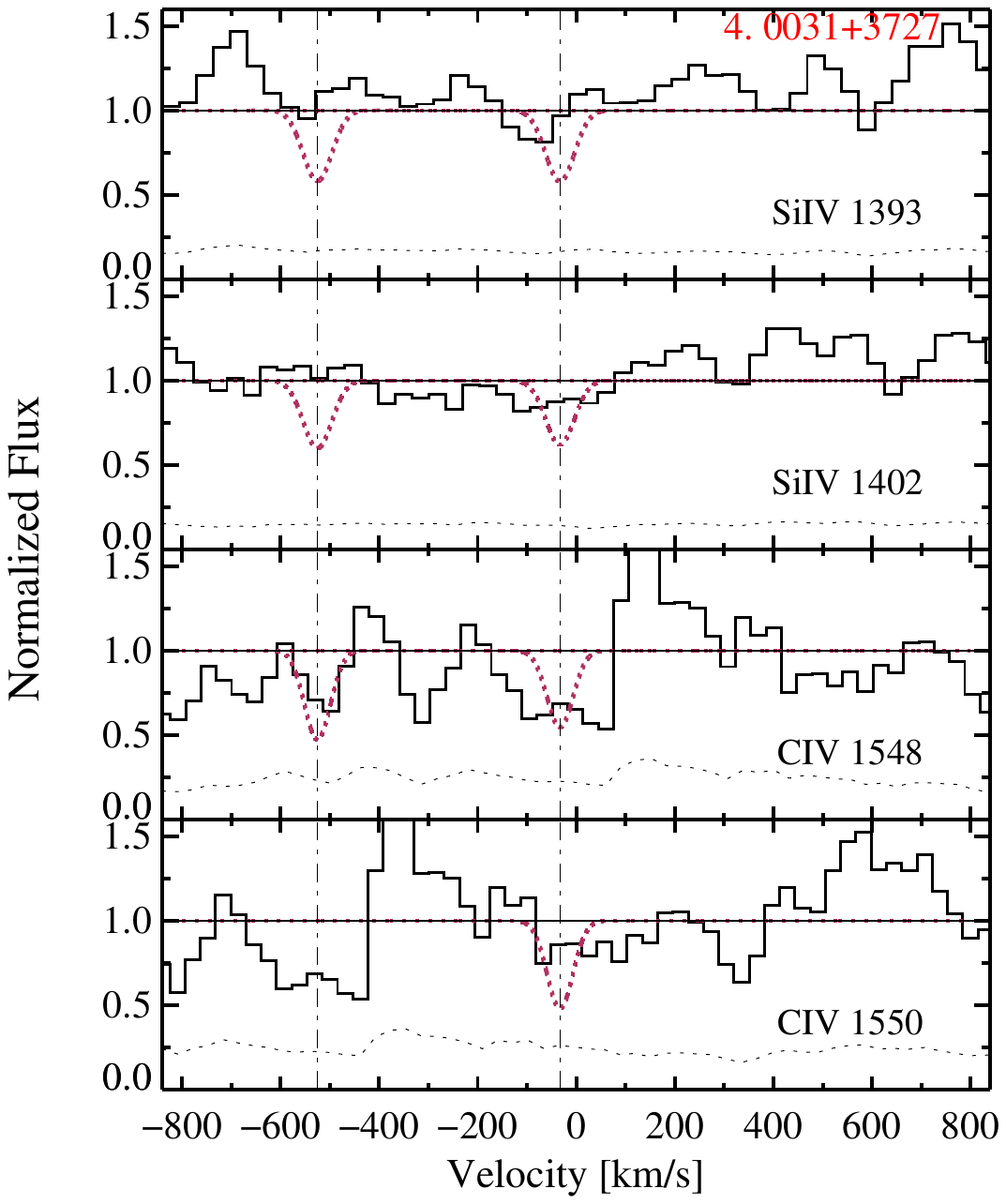}
\caption{Same as Figure 3, but for 0031+3727.}
\end{figure*}

\item[{\bf 5. 0032+3946 ($b=31.5$ kpc, $\lambda_{forest} < 2600$ \AA,
Fig. 7):}] Only NUV spectra were obtained for this quasar.  An M31
\MgII\ $\lambda2796$ absorption line with a significance of $2\sigma$
at a heliocentric velocity  of $-453$ km s$^{-1}$ appears to be
present (see Table 3), however a corresponding 2-pixel-wide
absorption feature near the expected position of \MgII\ $\lambda2803$
has a significance $<2\sigma$. If present, this absorption may
originate at the  southwest edge of M31's disk (see Figure 1). Apart
from strong Galactic emission,  the GBT 21 cm data along this
sightline (bottom panel of Figure 7) shows evidence for M31 emission
between $-509$ and $-459$ km s$^{-1}$.  Although the resolution of the
NUV spectrum is $\sim0.82$ \AA\ ($\sim 87$ km s$^{-1}$) at the
position of \MgII, the centroid of the absorption line can be
estimated with an uncertainty of $\sim 6$ km s$^{-1}$ (see \S4). Thus,
the identified \MgII\ $\lambda2796$ feature at $-453$ km s$^{-1}$   is
near the maximum velocity of observed 21 cm emission. Keeping in mind
the limitations of using \HI\ 21 cm emission observations to determine
\HI\ column  densities (\S2.1), we find $N_{HI} \approx 2.5\times
10^{18}$ atoms cm$^{-2}$ along this sightline. Very significant
Galactic \MgII\ and \FeII\ absorption is detected along this
sightline, but the Galactic \FeII\ $\lambda2586$ line was deblended to
separate it from unrelated  nearby absorption.

\begin{figure*}
\vspace{1.0in} 
\includegraphics[width=2.5in]{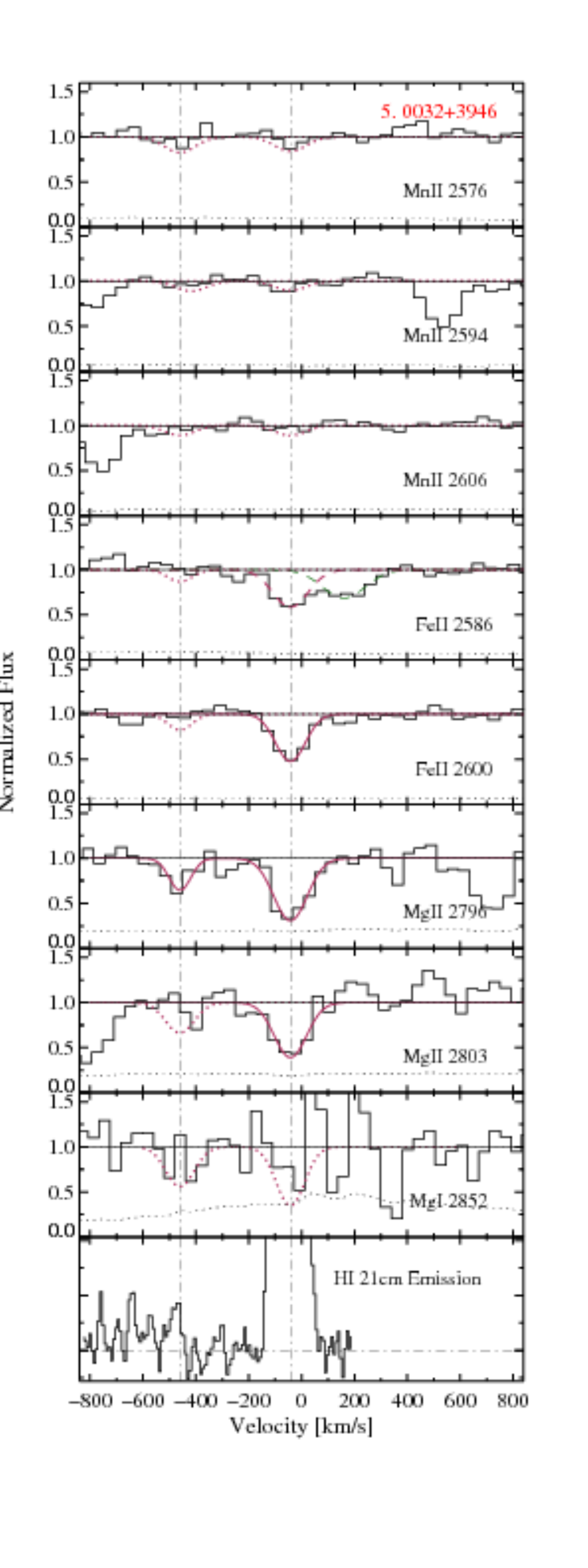}
\caption{Same as Figure 3, but for 0032+3946. In addition, the
\HI\ 21 cm emission profile extracted from the GBT data of Thilker et
al. (2004) is shown in the bottom panel.  The intensities are scaled 
to accentuate the very weak emission signal from M31.  The
dash-dot horizontal line drawn in the 21 cm panel marks the location
of zero intensity. No FUV spectra of this quasar were
obtained.}
\end{figure*}

\item[{\bf 6. 0037+3908 ($b=30.5$ kpc, $\lambda_{forest} < 2590$ \AA,
Fig. 8):}] Only NUV spectra were obtained for this quasar.
Absorption from M31 gas is not detected. However, apart from the
strong Galactic emission,  the GBT data  along this sightline reveal
M31 21 cm emission between $-542$ and $-475$ km s$^{-1}$ (bottom panel
of Figure 8), with an integrated column density of $N_{HI}=2.5\times
10^{18}$ atoms cm$^{-2}$ (see \S2.1).  The $2\sigma$ upper limits on
M31 absorption are made at the central velocities predicted by the
observed M31 21 cm emission.  Very significant Galactic \MgII\ and
\FeII\ absorption is detected along this sightline. The Galactic
\FeII\ $\lambda2586$ and \FeII\ $\lambda2600$ lines were deblended  to
separate them out from unrelated nearby absorption.

\begin{figure*}
\vspace{1.0in}  
\includegraphics[width=2.5in]{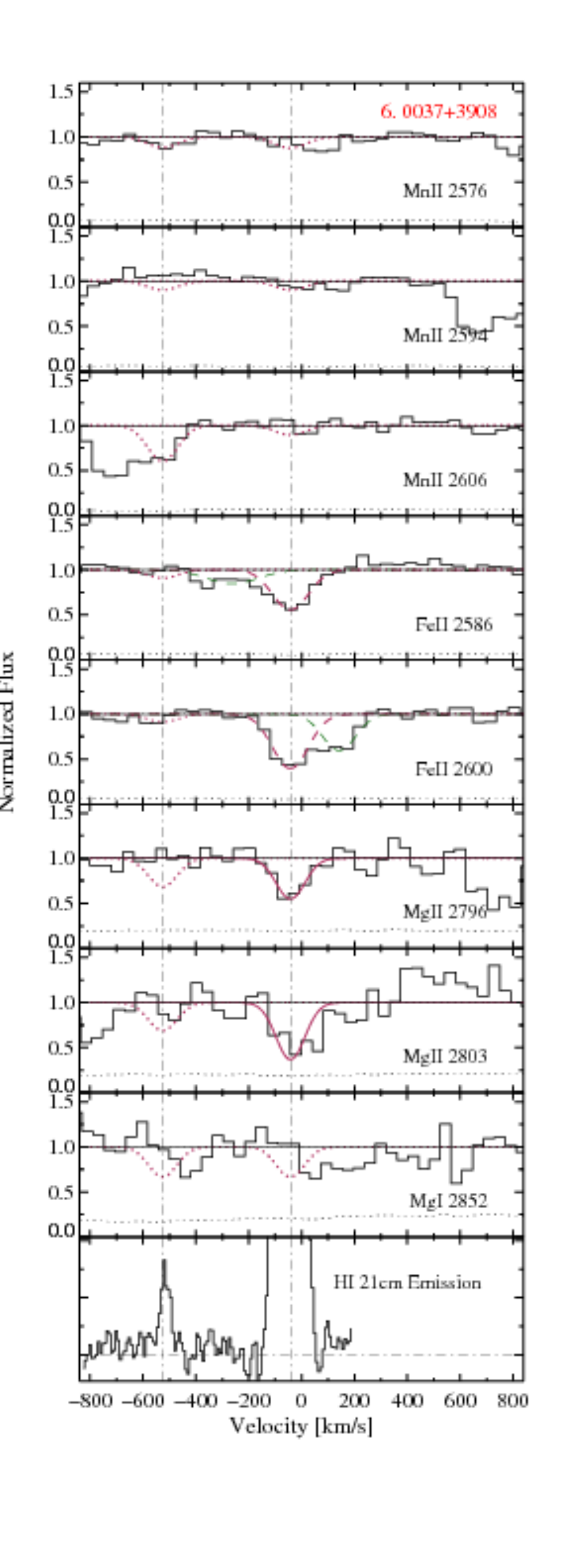}
\caption{Same as Figure 7, but for 0037+3908. No FUV spectra were
obtained for this quasar.}
\end{figure*}

\item[{\bf 7. 0040+3915 ($b=26.9$ kpc, $\lambda_{forest} < 2552$ \AA,
Fig. 9):}]  Only the velocity profiles in the vicinity of \MgII\ and
\MgI\  are visible in our observations for two reasons.  First, the
quasar spectrum exhibits intrinsic broad absorption lines (BALs) and
the \NV\ BAL  trough overlaps the \MnII\ and \FeII\ absorption-line
regions. This prevents useful measurements  of M31 and Galactic lines
in those regions.  Second, the FUV spectrum shows no useful continuum
flux, possibly due to strong  shorter-wavelength BALs and/or
overlapping intervening Lyman limit absorption.  \MgII\ $\lambda2796$
due to M31 gas appears as two absorption  components in the NUV
spectrum. The noise characteristics of the spectrum are worse  in the
corresponding \MgII\ $\lambda2803$ region, and two absorption
components are not seen (a single Gaussian was fitted to the
absorption),  but we give this lower weight due to the higher noise.
The two vertical dash-dot lines at $-389$ km s$^{-1}$ and $-513$ km
s$^{-1}$ mark the velocity positions of the two  M31 \MgII\
$\lambda2796$ absorption components.  The sightline passes through an
HVC (see Figure 1) , whose 21 cm emission profile can clearly be seen
in the bottom panel of the figure peaking at $\sim -500$ km
s$^{-1}$. The GBT data reveal that this 21 cm  emission extends
between $-542$ and $-442$ km s$^{-1}$.  Thus, the two \MgII\
absorption components at $-389$ km s$^{-1}$ and $-513$ km s$^{-1}$ may
correspond to M31 halo gas and HVC gas, respectively, with the halo
component showing no apparent 21 cm emission.  From the WSRT 21 cm
emission data, the integrated \HI\ column density in the HVC is
estimated to be $N_{HI}=9.5\times 10^{19}$ atoms cm$^{-2}$ (see
\S2.1). Very significant Galactic \MgII\ absorption is present along
this sightline.

\begin{figure*}
\vspace{-3.0in}  
\includegraphics[width=3.0in]{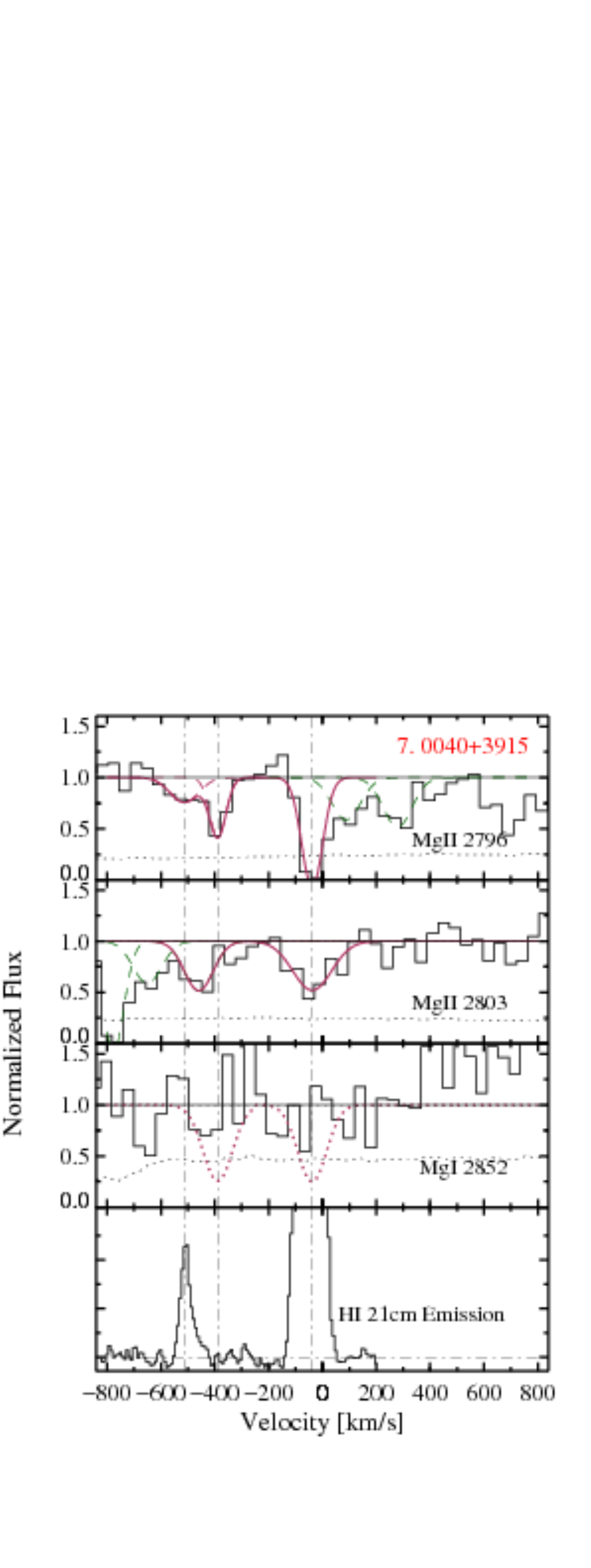}
\caption{Same as Figure 7, but for 0040+3915. The M31 HVC that is
detected  in 21 cm at $-513$ km/s, is also detected in the
\MgII$\lambda2796$ line.  The two \MgII$\lambda2803$ components are
too weak to be resolved with  these data. The FUV data are not shown
because the spectrum had no flux presumably due to an intervening
Lyman limit system.}
\end{figure*}

\item[{\bf 8. 0043+4016 ($b=13.4$ kpc, $\lambda_{forest} < 2545$ \AA,
Fig. 10):}]  This is the lowest impact parameter sightline. M31
low-ion absorption from \SiII\ $\lambda1260$ and \CII\ $\lambda1334$
is detected, and high-ion absorption from  \CIV\ $\lambda1548$ is
detected, but the \SiII\ $\lambda1260$ and \CIV\ $\lambda1548$ lines
had to be deblended from overlapping unrelated absorption. Given that
the 21 cm emission extends over a large range in velocity, we cannot
rule out that all the absorption features within the \CIV\ $\lambda1548$ blend
are due to \CIV\ $\lambda1548$ absorption over a wide velocity range. Confirmation
would require a higher signal-to-noise spectrum; here, we identify the 
lowest velocity component with the M31 \CIV\ $\lambda1548$ absorption line.
The low-ions are centered at $\sim -336$ km s$^{-1}$ and the high-ions are
centered at $\sim -340$ km s$^{-1}$. However, \MgII\ and \FeII\
absorption lines from M31 gas were not detected.  The GBT data show
that 21 cm emission from M31 exists along this sightline between
$-559$ km s$^{-1}$ and $-326$ km s$^{-1}$, with a total integrated
column density of $N_{HI}=1.2\times 10^{19}$ atoms cm$^{-2}$. We note
that the absorption-line velocities are coincident with the peak in
the 21 cm emission-line spectrum (bottom panels of Figure 10).
Many significant Galactic absorption lines are present. Galactic
\SiII\ $\lambda1260$ and  \CIV\ $\lambda1550$ had to be deblended to
separate them out from unrelated overlapping absorption.
 
\begin{figure*}
\vspace{1.0in}
\includegraphics[width=2.2in]{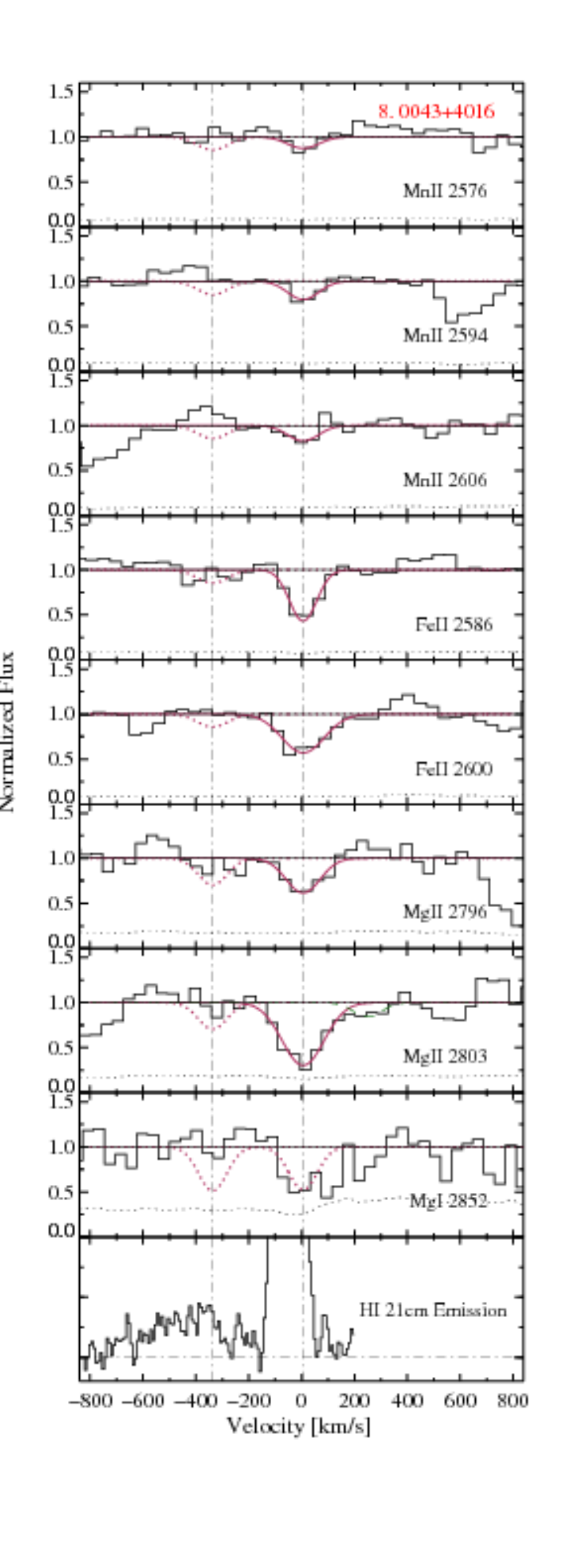}
\includegraphics[width=2.2in]{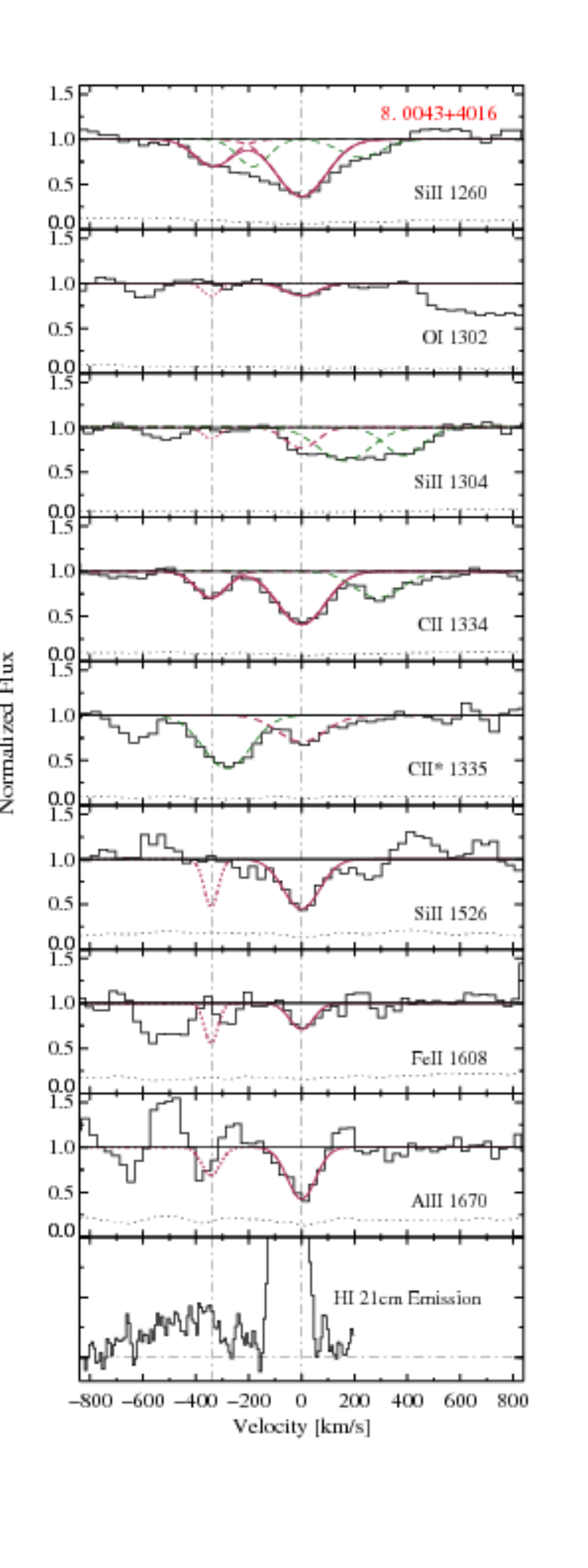}
\includegraphics[width=2.2in]{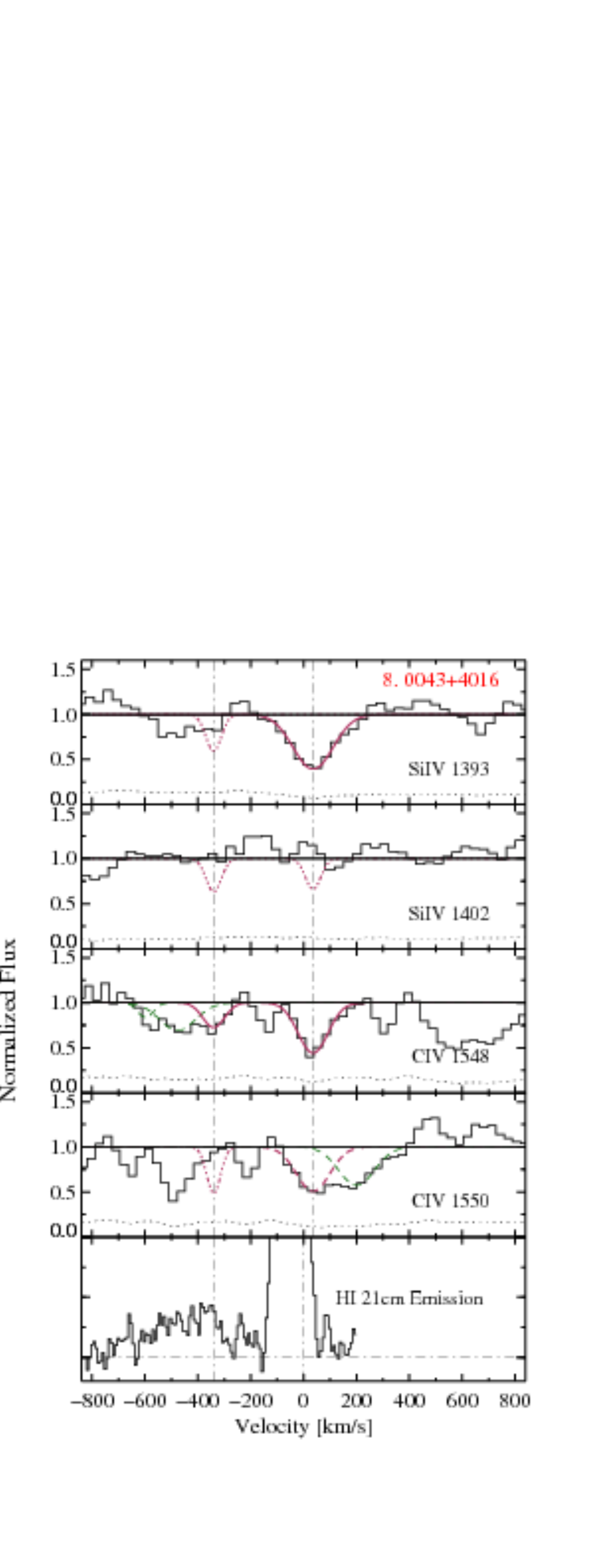}
\caption{Same as Figure 7, but for 0043+4016. FUV data obtained for this
quasar are also shown. }
\end{figure*}

\item[{\bf 9. 0043+4234 ($b=17.4$ kpc, $\lambda_{forest} < 1448$ \AA,
Fig. 11):}]  The sightline to this quasar passes ``above'' the \HI\
21 cm emission disk  of M31 on the receding, northwest, side (see
Figure 1).  Due to the location of the sightline, the detected M31 and
Galactic absorption lines needed to be deblended from each other.  We
used two-component Gaussian fits with fixed velocity components to do
this. Measurements of the  \SiII\ $\lambda$1260, \CII\ $\lambda$1334.5
(and \CIIs\ $\lambda$1335.7), \SiIV\ $\lambda1393$, and \AlII\ $\lambda1670$  absorption
lines are further complicated by other overlapping absorption.  For
low-ion absorption the velocity centroid for the Galactic lines was
fixed using \MnII\ $\lambda2576$, while allowing the position of the
M31 low-ion velocity centroid to vary until the best least-squares
solution was found. Deblending indicates that the detected  M31
low-ion gas, which gives rise to transitions of \SiII, \CII, \AlII,
\FeII\ and \MgII, is   located at $-234$  km s$^{-1}$, and the
Galactic low-ion gas is located at  $-73$ km s$^{-1}$. It is notable
that along this sightline there is a  detection of Galactic \CIIs\
$\lambda$1335.7.  The M31 high-ion gas, which gives rise to
transitions of \SiIV\ $\lambda$1393 and  \CIV\ $\lambda$1548, are also
members of a multi-component blend  with Galactic lines. Using a
procedure similar to the one used for the low-ions, we find that the
M31 high-ion gas is at $-191$ km s$^{-1}$ and the Galactic high-ion
gas  is at $-1$ km s$^{-1}$.  GBT data reveal M31 21 cm emission
between $-259$ km s$^{-1}$ and $-93$ km s$^{-1}$, with a total
integrated column density of  $N_{HI}=8\times10^{18}$ cm$^{-2}$ (see
\S2.1). The higher velocity edge of the M31 21 cm emission is
uncertain  since it may overlap with Galactic 21 cm emission.

\begin{figure*}
\vspace{1.0in}
\includegraphics[width=2.2in]{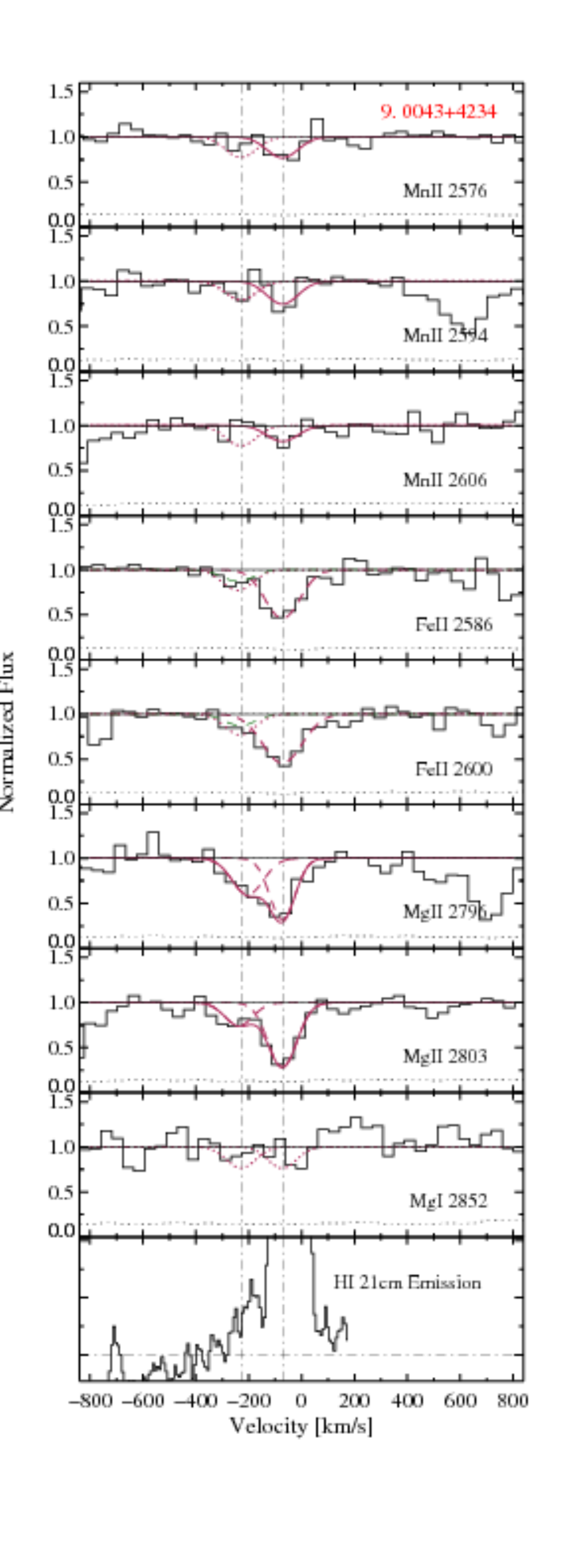}
\includegraphics[width=2.2in]{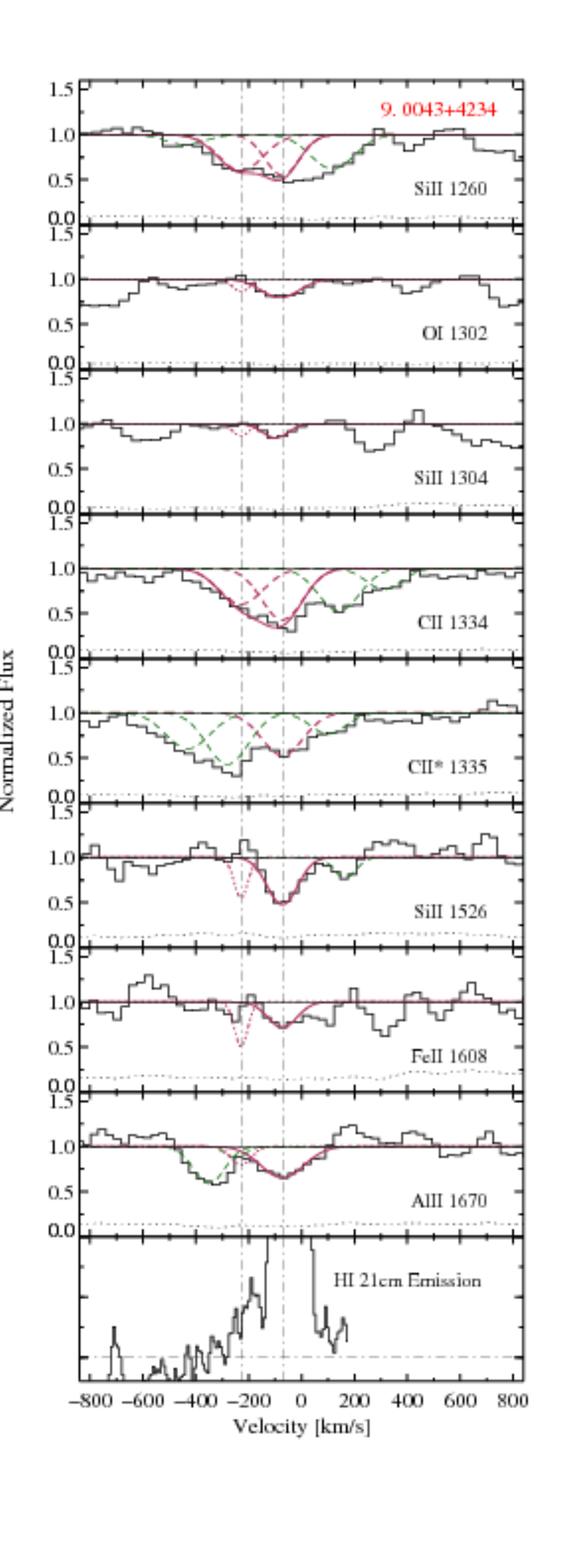}
\includegraphics[width=2.2in]{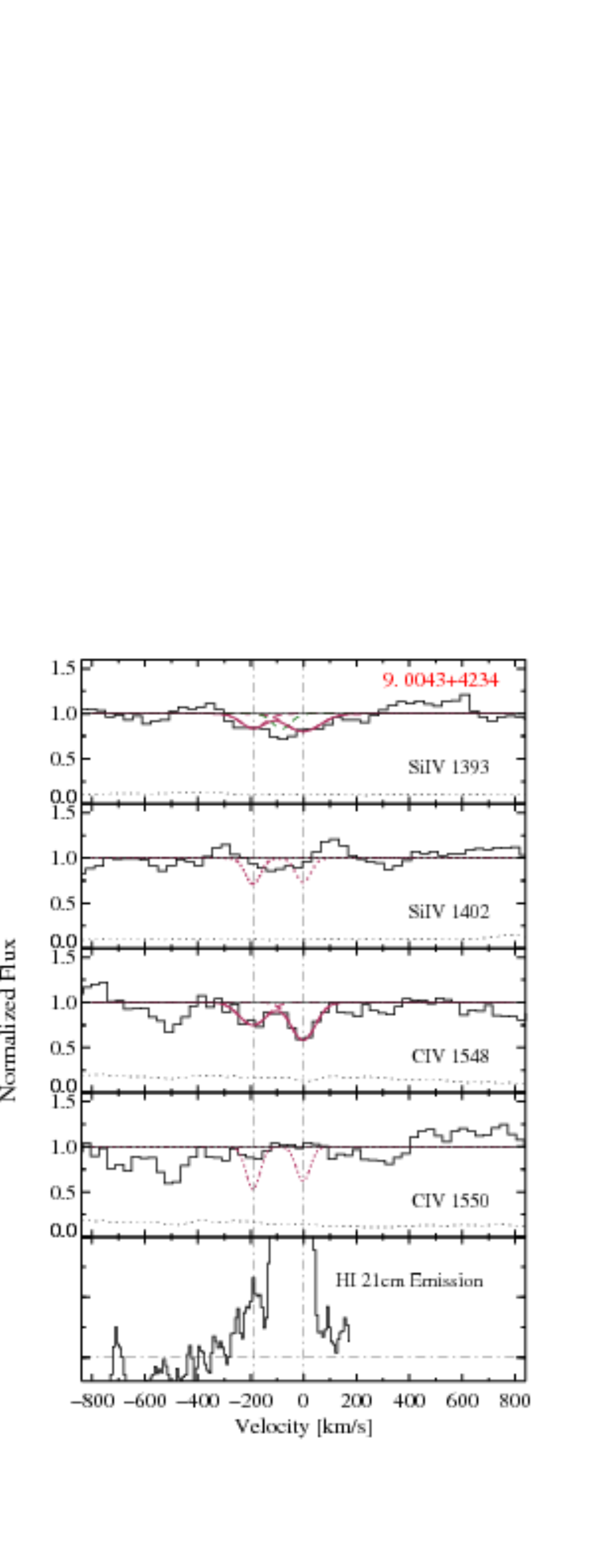}
\caption{Same as Figure 7, but for 0043+4234. FUV data obtained for this
quasar are also shown.}
\end{figure*}

\item[{\bf 10. 0046+4220 ($b=17.5$ kpc, $\lambda_{forest} < 1588$ \AA,
Fig. 12):}]

To infer what gaseous structures exist along this
sightline we are guided by the observed GBT \HI\ 21 cm emission
velocity profile, which is shown in the bottom panels of Figure 12.
An inset in the bottom left panel shows the entire 21 cm profile. Most
notably, the weaker 21 cm peak near $-5$ km s$^{-1}$ represents
Galactic emission, while the stronger 21 cm emission peak near $-55$
km s$^{-1}$ represents M31's disk; however, such a velocity
separation cannot be distinguished in the COS G140L and G230L
absorption-line spectra. More generally, the entire 21 cm velocity
profile and the detected low-ion absorption lines have allowed us to
infer the existence of five gaseous structures along this sightline:
one near $-5$ km s$^{-1}$ (Galactic gas), one near $-55$ km s$^{-1}$
(M31 disk gas), one near $-195$ km s$^{-1}$ (M31 halo gas), one near
$-334$ km s$^{-1}$ (a M31 HVC), and one near $-409$ km s$^{-1}$ (a
second M31 HVC).  As with sightline 9, detected Galactic and M31
absorption lines needed to be deblended from each other, but the
blending along this sightline is more severe. In particular, the
low-ion absorption detected near $-40$ km s$^{-1}$ must be a blend of
Galactic gas and M31 disk gas, with most of the absorption being due
to M31 disk gas.  This Galactic+M31 blended component is included 
under the heading of ''Milky Way Absorption Lines'' in Table 3 but 
with a footnote. Absorption due to \CIIs\ $\lambda$1335.7 is among the
many transitions detected in this component (see Table 3). A weak
(barely significant) blended Galactic and M31 high-ion absorption
component is located near $-35$ km s$^{-1}$.  The 21 cm emission
profile allows us to estimate that the Galactic component peaking near
$-5$ km s$^{-1}$ has $N_{HI}=4.0\times10^{20}$ atoms cm$^{-2}$ and the
M31 disk component peaking near $-55$ km s$^{-1}$ has
$N_{HI}=1.5\times10^{21}$ atoms cm$^{-2}$. Aside from this first
blended absorption component, a second low-ion absorption component is
seen displaced toward lower velocities by $\sim 155$ km s$^{-1}$,
close to the edge of the \HI\ 21 cm emission profile, which we take as
evidence for halo gas. However, measurements of the \SiII\ $\lambda$1260, 
\OI\ $\lambda$1302, \SiII\ $\lambda$1304, \CII\
$\lambda$1334.5 (and \CIIs\ $\lambda$1335.7), \SiII\ $\lambda1526$,
\CIV\ $\lambda1548$, and \AlII\ $\lambda1670$ absorption lines are
complicated by overlapping or nearby absorption.  Deblending indicates
that the M31 low-ion halo gas component is near $-195$ km s$^{-1}$,
and this gives rise to absorption due to \SiII, \OI, \CII, \AlII,
\FeII, and \MgII.  Deblending also indicates that a high-ion
absorption component is located near $-152$ km s$^{-1}$; it is clearly
present in \CIV\ but possibly not \SiIV.  The 21 cm emission allows us
to estimate that the M31 halo component peaking near $-195$ km
s$^{-1}$ has $N_{HI}=7\times10^{19}$ atoms cm$^{-2}$. The velocity
locations of the above described absorption components for the low
ions and high ions are shown as vertical dot-dashed lines in the
panels, including the lower left inset panel. In addition, the GBT 21
cm observations also reveal gas from two M31 HVCs near $-334$ km
s$^{-1}$ (between $-342$ km s$^{-1}$ and $-326$ km s$^{-1}$), and near
$-409$ km s$^{-1}$ (between $-426$ km s$^{-1}$ and $-392$ km
s$^{-1}$).  The total integrated column densities along the sightlines
to these HVCs are $N_{HI}=2\times10^{18}$ atoms cm$^{-2}$ and
$N_{HI}=2.5\times10^{18}$ atoms cm$^{-2}$,  respectively. We do not
detect metal-line absorption near the velocities of these HVCs, so we
have not used vertical lines to mark their velocity locations in
Figure 12. Thus, using standard quasar absorption line jargon, we
conclude that, given the estimated $N_{HI}$ values for the four
detected M31 velocity components, we have detected a DLA system (M31
disk gas), a sub-DLA system (M31 halo gas), and two Lyman limit
systems (two M31 HVCs).

Finally, we point out that the blended low-ion absorption near $-40$
km s$^{-1}$ in sightline 10 is the only system which reaches DLA \HI\
column densities (i.e., $N_{HI} \ge 2\times 10^{20}$ atoms cm$^{-2}$).
As described above, it is due to a blend of Galactic gas and M31 disk
gas. DLAs are the quasar absorption-line systems used to track the
evolution of neutral gas in the Universe at low (Rao et al. 2006) and
high (e.g., Noterdaeme et al. 2012) redshift. The strength of the
\MgII\ $\lambda2796$ and \FeII\ $\lambda2600$ absorption lines in this
component are consistent with criteria used in \MgII-selected DLA
searches (Rao et al. 2006).

\begin{figure*}
\vspace{1.0in}
\includegraphics[width=3.5in]{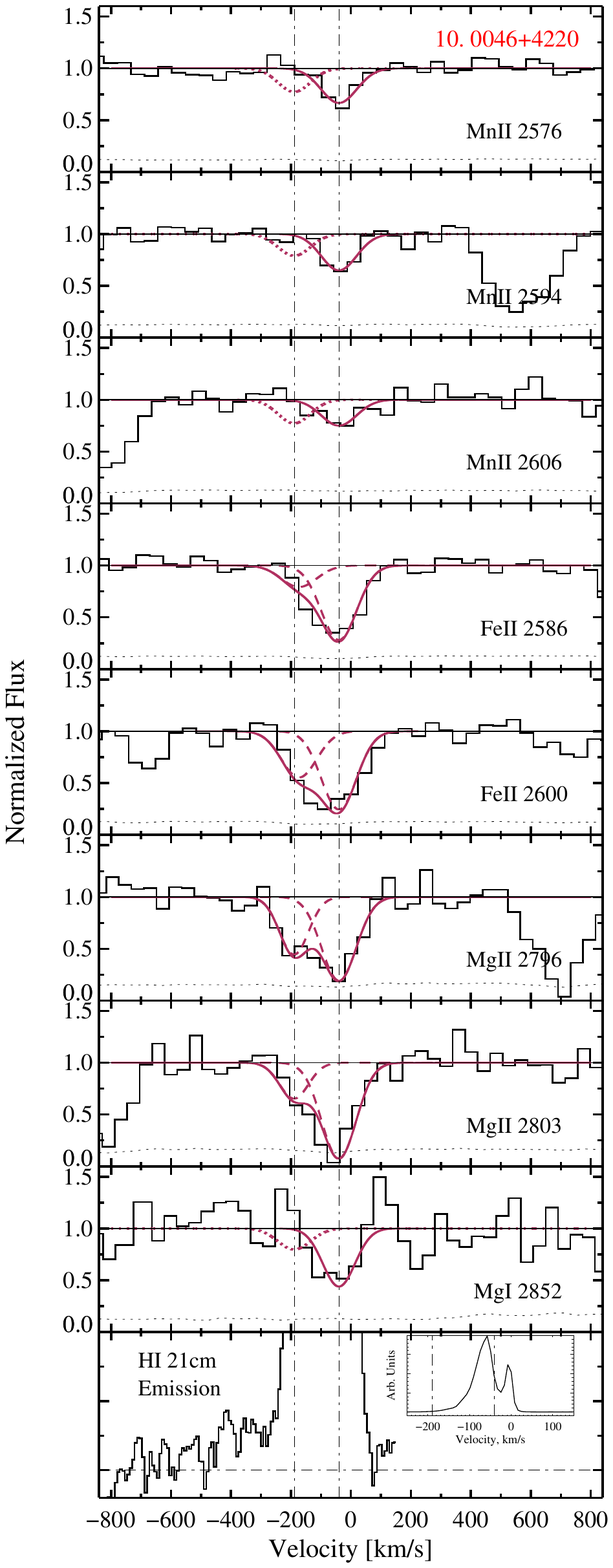}\hspace{-1.6in}
\includegraphics[width=3.5in]{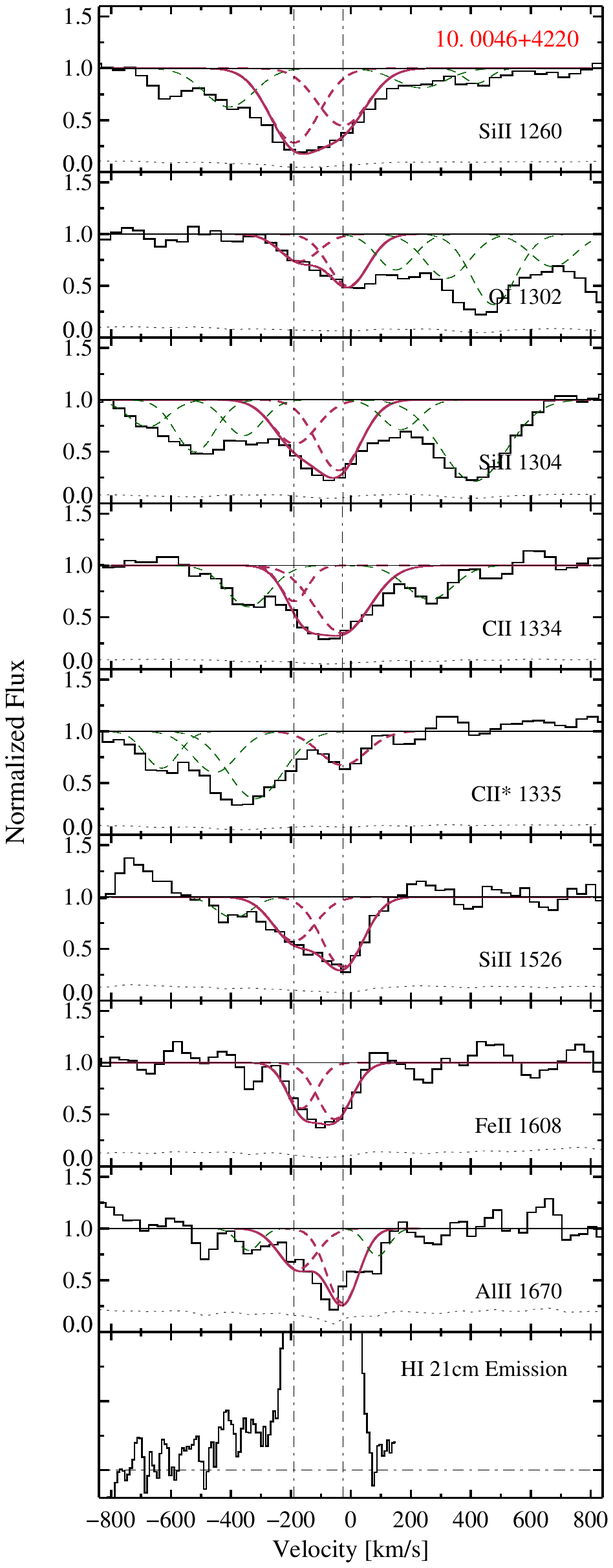}\hspace{-1.8in}
\includegraphics[width=3.3in]{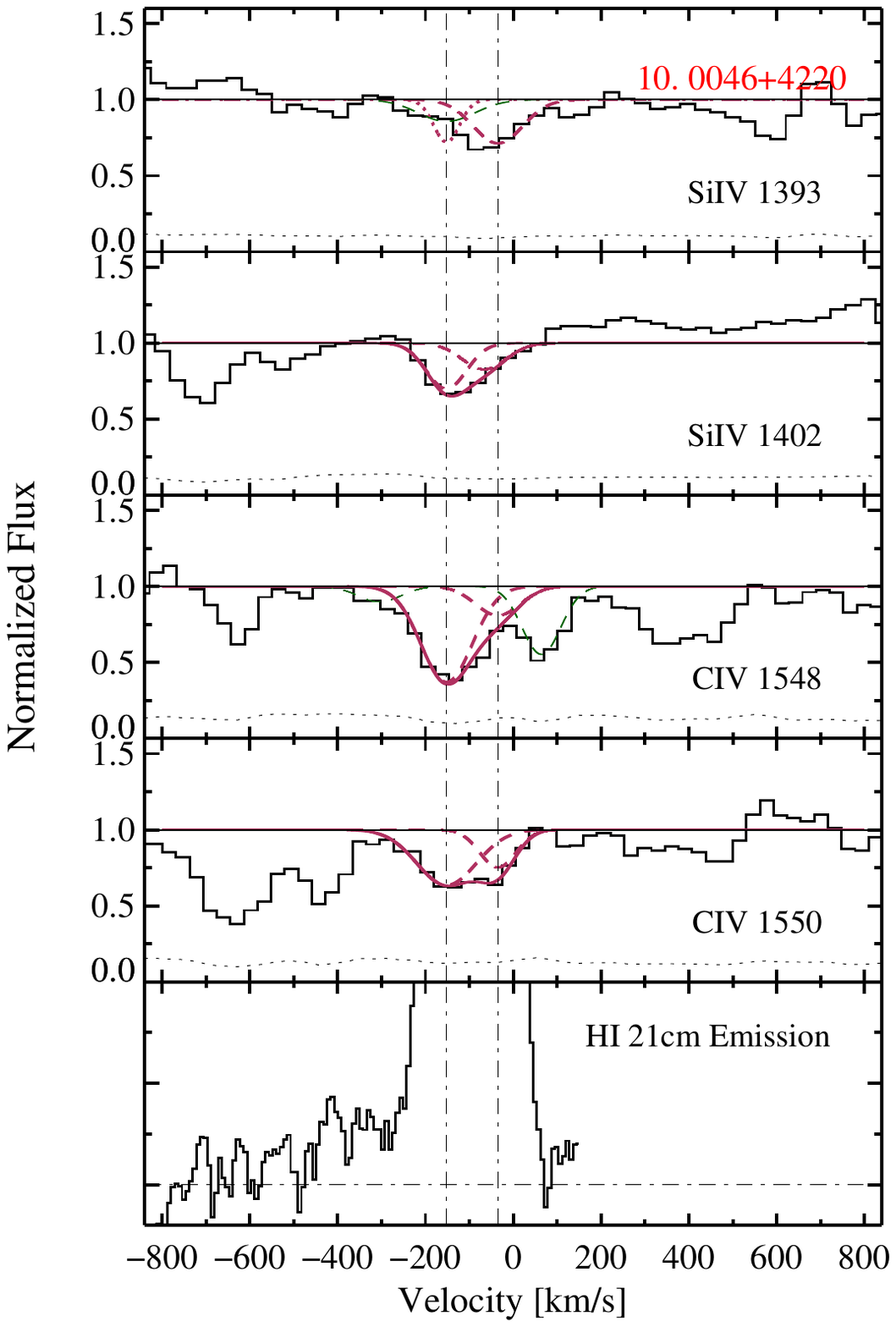}
\caption{Same as Figure 7, but for 0046+4220. FUV data obtained for this
quasar are also shown. }
\end{figure*}

\end{description}

\begin{table*}
\hspace*{-1.0in}
\tiny
\caption{Rest Equivalent Width Measurements\tablenotemark{a}}
\begin{tabular}{@{}lcccccccccc}
\hline
\hline
Line  & 1.0018+3412 & 2.0024+3439 & 3.0030+3700 & 4.0031+3727 & 5.0032+3946 & 
6.0037+3908 & 7.0040+3915 & 8.0043+4016 & 9.0043+4234 & 10.0046+4220\\
 & REW (\AA) & REW (\AA) & REW (\AA) & REW (\AA)& REW (\AA)& REW (\AA) & REW (\AA)& REW (\AA)& REW (\AA)& REW (\AA)\tablenotemark{b}\\
\hline
\multicolumn{11}{c}{\bf M31 Absorption Lines}\\
\hline
SiII1260&     $\le0.113$      &        \nodata &       $\le0.300$      &       $\le0.132$      &       \nodata &       \nodata &       \nodata &       $0.225\pm0.042$ &       $0.259\pm0.054$         &       $0.544\pm0.042$\\
OI1302 &      $\le0.113$      &        \nodata &       $\le0.093$      &       $\le0.065$      &       \nodata &       \nodata &       \nodata &       $\le0.048$      &       $\le0.048$              &       $0.187\pm0.047$\\
SiII1304&     \nodata         &        \nodata &       \nodata         &       \nodata         &       \nodata &       \nodata &       \nodata &       $\le0.043$      &       $\le0.046$              &       $0.341\pm0.036$\\
CII1334&	$\le0.106$	&	\nodata	&	$\le0.303$	&	$\le0.121$	&	\nodata	&	\nodata	&	\nodata	&	$0.192\pm0.042$ &	$0.318\pm0.039$	        &	$0.234\pm0.044$\\
CII$^*$1335&     $\le0.131$	&	\nodata	&	$\le0.373$	&	$\le0.118$	&	\nodata	&	\nodata	&	\nodata	&	\nodata	        &	\nodata	                &	\nodata\\
SiIV1393&	$\le0.145$	&	\nodata	&	$\le0.205$	&	$\le0.143$	&	\nodata	&	\nodata	&	\nodata	&	$\le0.139$      &	$0.085\pm0.042$		&	$\le0.095$\\
SiIV1402&	$\le0.131$	&	\nodata	&	$\le0.188$	&	$\le0.137$	&	\nodata	&	\nodata	&	\nodata	&	$\le0.125$	&	$\le0.098$		&	$0.156\pm0.065$	\\
SiII1526&	$\le0.173$	&	\nodata	&	$\le0.523$	&	$\le0.214$	&	\nodata	&	\nodata	&	\nodata	&	$\le0.180$	&	$\le0.152$		&	$0.386\pm0.066$	\\
FeII1608&	$\le0.223$	&	\nodata	&	$\le0.384$	&	$\le0.209$	&	\nodata	&	\nodata	&	\nodata	&	$\le0.151$	&	$\le0.168$		&	$0.267\pm0.076$	\\
AlII1670&	$\le0.244$	&	\nodata	&	$\le0.453$	&	$\le0.253$	&	\nodata	&	\nodata	&	\nodata	&	$\le0.194$	&	$\le0.110$		&	$0.258\pm0.087$	\\
CIV1548&	$\le0.202$	&	\nodata	&	$\le0.298$	&	$\le0.180$	&	\nodata	&	\nodata	&	\nodata	&	$0.146\pm0.061$ &	$0.167\pm0.087$	        &	$0.651\pm0.098$	\\
CIV1550&	\nodata        &	\nodata	&	\nodata 	&	\nodata 	&	\nodata	&	\nodata	&	\nodata	&	$\le0.175$	&	$\le0.161$		&	$0.320\pm0.070$	\\
MnII2576&	$\le0.447$	&   $\le0.274$	&	$\le0.307$	&	$\le0.829$	&    $\le0.218$ &   $\le0.150$ &	\nodata	&	$\le0.181$	&	$\le0.285$		&	$\le0.243$	\\
MnII2594&	$\le0.422$	&   $\le0.283$	&	$\le0.302$	&	$\le0.815$	&    $\le0.139$ &   $\le0.123$ &	\nodata	&	$\le0.187$	&	$\le0.263$		&	$\le0.257$	\\
MnII2606&	$\le0.428$	&   $\le0.291$	&	$\le0.311$	&	$\le0.829$	&    $\le0.138$ &   $\le0.498$ &	\nodata	&	$\le0.182$	&	$\le0.285$		&	$\le0.259$	\\
FeII2586&	$\le0.421$	&   $\le0.254$	&	$\le0.435$	&	$\le0.847$	&    $\le0.178$ &   $\le0.120$ &	\nodata	&	$\le0.187$	&	$\le0.289$		& 	$ 0.266\pm0.115$\\
FeII2600&	$\le0.430$	&   $\le0.291$	&	$\le0.313$	&	$\le0.865$	&    $\le0.133$ &   $\le0.114$ &	\nodata	&	$\le0.183$	&	$\le0.291$		&	$ 0.571\pm0.112$\\
MgII2796&       $\le0.459$	&   $\le0.261$	&	$\le0.211$	&	$\le0.433$	&$0.341\pm0.171$&   $\le0.405$ &  $0.304\pm0.220$\tablenotemark{c}  &$\le0.379$    &$0.554\pm0.111$     &       $0.627\pm0.112$\\
\nodata	&       \nodata	        &       \nodata	&        \nodata	&        \nodata	&       \nodata	&      \nodata &  $0.412\pm0.256$\tablenotemark{c}      &\nodata &\nodata               & \nodata\\
MgII2803&       $\le0.484$	&   $\le0.248$	&	$\le0.226$	&	$ \le0.354$	&    $\le0.430$ &   $\le0.399$ &  $0.606\pm0.208$	&	$\le0.377$		&	$ 0.285\pm0.117$ & $ 0.390\pm0.122$\\
MgI2852&	$\le0.619$	&   $\le0.387$	&	$\le0.419$	&	\nodata		&    $\le0.565$ &   $\le0.428$ &  $\le0.981$    &   $\le0.624$	&	$\le0.301$		&	$\le 0.260$\\
\hline
\multicolumn{11}{c}{\bf Milky Way Absorption Lines}\\
\hline\\
SiII1260&     $0.289\pm0.054$ &     \nodata	&	$0.358\pm0.081$	&	$0.320\pm0.053$	&	\nodata	&	\nodata	&	\nodata	&	$0.636\pm0.048$	&	$0.689\pm0.044$	&	$0.479\pm0.056$\\
OI1302 &      $0.189\pm0.086$ &     \nodata    &       $\le0.071$      &       $0.113\pm0.042$ &       \nodata &       \nodata &       \nodata &       $0.107\pm0.021$ &       $0.162\pm0.032$ &       $0.358\pm0.050$\\
SiII1304&     $0.258\pm0.033$ &     \nodata    &       $\le0.065$      &       $\le0.050$      &       \nodata &       \nodata &       \nodata &       $0.154\pm0.029$ &       $0.084\pm0.032$ &       $0.545\pm0.045$\\
CII 1334&      $0.410\pm0.095$ &     \nodata	&	$0.448\pm0.128$	&	$0.309\pm0.048$	&	\nodata	&	\nodata	&	\nodata	&	$0.594\pm0.034$	&	$0.592\pm0.036$	&	$0.669\pm0.032$	\\
CII$^*$1335&     $\le0.129$      &     \nodata	&	$\le0.401$	&	$\le0.115$	&	\nodata	&	\nodata	&	\nodata	&	$0.295\pm0.044$	&	$0.413\pm0.044$	&	$0.288\pm0.039$	\\
SiIV1393&	$\le0.131$     &     \nodata	&	$0.396\pm0.121$	&	$\le0.144$	&	\nodata	&	\nodata	&	\nodata	&	$0.533\pm0.036$	&	$0.149\pm0.056$	&	$0.189\pm0.095$	\\
SiIV1402&	$\le0.108$     &     \nodata	&	$\le0.163$	&	$\le0.130$	&	\nodata	&	\nodata	&	\nodata	&	$\le0.116$	&	$\le0.104$	&	$0.133\pm0.065$	\\
SiII1526&     $0.286\pm0.046$ &     \nodata	&	$\le0.462$	&	$0.155\pm0.097$	&	\nodata	&	\nodata	&	\nodata	&	$0.483\pm0.062$	&	$0.706\pm0.093$	&	$0.761\pm0.092$	\\
FeII1608&	$\le0.231$     &     \nodata	&	$\le0.372$	&	$0.243\pm0.092$	&	\nodata	&	\nodata	&	\nodata	&	$0.175\pm0.076$	&	$0.119\pm0.060$	&	$0.424\pm0.078$	\\
AlII1670&     $0.634\pm0.063$ &     \nodata	&	$\le0.557$	&	$0.443\pm0.102$	&	\nodata	&	\nodata	&	\nodata	&	$0.427\pm0.082$	&	$0.274\pm0.066$	&	$0.440\pm0.099$	\\
CIV1548&      $0.213\pm0.077$ &     \nodata	&	$0.471\pm0.261$	&	$\le0.155$	&	\nodata	&	\nodata	&	\nodata	&	$0.270\pm0.110$	&	$0.252\pm0.078$	&	$0.264\pm0.079$	\\
CIV1550&      $0.365\pm0.074$ &     \nodata	&	$\le0.299$	&	$\le0.177$	&	\nodata	&	\nodata	&	\nodata	&	$0.378\pm0.095$	&	$\le0.128$	&	$0.132\pm0.066$	\\
MnII2576&	$\le0.417$     &$0.356\pm0.106$	&	$\le0.722$	&	$\le0.849$	&   $\le0.205$	&   $\le0.150$	&	\nodata	&	$0.135\pm0.067$ &	$0.297\pm0.112$	&	$0.427\pm0.093$	\\
MnII2594&	$\le0.428$     & $\le0.285$	&	$\le0.317$	&	$\le0.838$	&   $\le0.133$	&   $\le0.123$	&	\nodata	&	$0.222\pm0.074$	&	$0.325\pm0.104$	&	$0.448\pm0.100$	\\
MnII2606&	$\le0.438$     & $\le0.273$	&	$\le0.310$	&	$\le0.884$	&   $\le0.143$	&   $\le0.136$	&	\nodata	&	$0.190\pm0.082$	&	$0.226\pm0.113$	&	$0.319\pm0.100$	\\
FeII2586&    $0.627\pm0.222$  &$0.626\pm0.126$	&	$0.667\pm0.163$	&	$\le0.838$	&$0.602\pm0.142$&$0.659\pm0.078$&	\nodata	&	$0.632\pm0.077$	&	$0.767\pm0.147$ &	$0.919\pm0.092$	\\
FeII2600&    $0.591\pm0.162$  &$0.676\pm0.139$	&	$0.671\pm0.172$	&	$1.209\pm0.312$	&$0.646\pm0.062$&$0.897\pm0.131$&	\nodata	&	$0.755\pm0.104$	&	$0.888\pm0.186$ &	$0.965\pm0.094$	\\
MgII2796&    $0.869\pm0.179$  &$0.968\pm0.146$	&	$1.058\pm0.115$	&	$0.726\pm0.133$	&$1.025\pm0.215$&$0.569\pm0.199$&$0.977\pm0.293$&       $0.580\pm0.194$ &	$0.764\pm0.107$ &	$1.096\pm0.122$	\\
MgII2803&    $0.508\pm0.198$  &$0.947\pm0.135$	&	$0.846\pm0.110$	&	$0.624\pm0.117$	&$0.870\pm0.220$&$1.038\pm0.248$&$0.089\pm0.288$&       $1.294\pm0.211$ &	$0.920\pm0.096$	&	$1.249\pm0.120$	\\
MgI2852&     $\le0.626$       &$\le0.493$	&	$\le0.595$	&	\nodata	        &$\le0.833$     &$\le0.428$     &$\le0.959$	&       $\le0.608$	&	$\le0.304$	&	$0.846\pm0.117$\\
\hline
\tablenotetext{a}{2$\sigma$ upper limits are tabulated for non-detections.} 
\tablenotetext{b}{In sightline 10, Milky Way absorption lines are blended with M31 disk gas. See the 
description in \S 3.}
\tablenotetext{c}{The two measurements are M31 HVC and disk components, respectively. See Figure 9.}
\end{tabular}
\hspace*{-1.0in}
\end{table*}

\section{Summary and Discussion of Results for M31}

\subsection{Overview on the Detection of Low-Ion and High-Ion Absorption Lines}

The detections of M31 gas presented in the previous section and
reported in  Tables 3 and 4 can be summarized as follows.

Low-ion \MgII\ absorption due to M31 gas is detected along four of the
10 observed sightlines (5, 7, 9, and 10). These sightlines
have impact parameters ranging between $b \approx$ 17 and 32 kpc.  We
also detect other low-ion gas (e.g., due to \SiII, \OI, \CII, \FeII,
or \AlII) along three of the four sightlines with \MgII\ detections;
sightline 5 was not observed in the FUV, where
most of these transitions occur.  In addition, we detect \CII\
absorption at M31 velocities along sightline 8 ($b=13.4$
kpc). Sightline 6 ($b=30.5$ kpc) is the only ``inner''
sightline which does not show evidence for M31 low-ion absorption
($W^{\lambda 2796}_0 < 0.41$ \AA);  however, no FUV spectra were
obtained along this sightline.   Among these ``inner'' sightlines, 
except for the blended Galactic and M31 line in sightline 10, the
\MgII\ rest equivalent widths ranged between $W^{\lambda 2796}_0$
$\approx$ 0.34 and 0.71 \AA, with the strongest detection being a
two-component absorber with $W^{\lambda 2796}_0$ $\approx$ 0.30 and
0.41 \AA. The four outer sightlines (1 through 4),  with
impact parameters $b \approx$ 57 to 112 kpc, do not show \MgII\
absorption down to $2\sigma$ rest equivalent upper limits ranging
between  $W^{\lambda 2796}_0$ $\approx$ 0.21 and 0.46 \AA.

High-ion \CIV\ absorption due to M31 gas is detected along three of
six sightlines (8, 9, and 10) which have usable FUV
spectra. These three detections are all in ``inner'' sightlines, with
impact parameters ranging between $b \approx$ 13 and 18 kpc, and rest
equivalent widths ranging between $W^{\lambda 1548}_0$ $\approx$ 0.17
and 0.65 \AA. Some \SiIV\ absorption and low-ion absorption is also
detected along these three ``inner'' sightlines.  The three \CIV\
non-detections are in outer sightlines (1, 3, and 4), with
impact parameters ranging between $b \approx$ 57 and 112 kpc, and with
$2\sigma$ rest equivalent width upper limits ranging between
$W^{\lambda 1548}_0$ $\approx$ 0.18 and 0.30 \AA.

We should point out that many of the detections summarized above were
near the limit of our  sensitivity threshold, despite the fact that
our rest equivalent width upper limits are typical of those in large
optical quasar absorption-line surveys. Another concern is confusion
from overlapping or nearby absorption, but we believe we have dealt
with this appropriately.

Also, Rich et al. (private communication) has observed three sightlines
in the halo of M31 with COS. They do not cover the \MgII\ region, 
but detect \CIV\ from M31 in some of these sightlines. There are other
HST archival observations in the M31 halo, but these do not show any detections. 

\subsection{Implications}

As elaborated further in \S4.3,  a clear picture does emerge. The
absorption lines that arise in M31 gas are found to be relatively weak
in comparison to those often identified in optical quasar
absorption-line  surveys, and even more so in comparison to absorption
lines which arise in the ISM of the Milky Way Galaxy (e.g., Table
3). Moreover, none of the detected M31 absorption lines are found at
large  impact parameters. This could also be viewed as unexpected
since the bulk of intervening  low- to moderate-redshift metal-line
absorbers seen in quasar spectra are identified with
large-impact-parameter galaxies in followup imaging studies (e.g.,
Rao et al. 2011, Chen et al. 2010).  However,
all of the large-impact-parameter sightlines we observed were
generally along M31's major axis, so one scenario which might explain
the lack of absorption in those cases would be to hypothesize that
extended gaseous absorption originates in  galactic fountains and
preferentially avoids extended regions along the direction of the
disk (e.g., Bordoloi et al. 2011, Bouch\'e et al. 2012).  Using the
observed distribution of HVCs around the Milky Way and M31, Richter (2012)
finds an exponential decline in the mean filling factor of HVCs with a 
characteristic radial extent of $\sim 50$ kpc. If HVCs alone are responsible
for absorption lines, then one would not expect to find any absorption 
along our four outer sightlines.
Alternatively, M31 may simply be typical of a class of
luminous galaxies that don't possess  large gaseous cross sections
which are capable of giving rise to moderate-strength quasar
absorption lines.   Our findings for M31 may in some way be connected
to the observed relative  decrease in the incidence of stronger \MgII\
systems with  decreasing redshift (e.g., Nestor et al. 2005).

In the past several years there has been speculation that M31 is a
galaxy that lies in the ``green valley¡Ç¡Ç (e.g., Mutch et al. 2011,
Davidge et al. 2012). The idea is that it exhibits properties that put
it between the red cloud and blue cloud populations that have been
identified in large galaxy surveys. Such galaxies may be in a stage of
transition and their star formation may nearly cease in less than 5
Gyrs. While this may be the case for M31, we note that the data we
have discussed here should not be taken to offer any clues about
this. For example, our data do not allow us to draw any conclusions
about the strength of star formation or even the column densities of
metal-line absorption. This is because the lines we have identified
are likely to be mostly saturated.  Thus, the weakness of the
metal-line absorption in M31 most likely indicates that the effective
gas velocity spread is low; it may either be truly low relative to the
spectral resolution and/or there may be a small number of velocity
components within the spectral resolution element.

\subsection{\MgII\ Rest Equivalent Width ($W_0^{\lambda2796}$) versus Impact Parameter ($b$)}

Figure 13 is a plot of M31 \MgII\ $\lambda2796$ rest equivalent width
($W_0^{\lambda2796}$) detections (or $2\sigma$ upper limits) versus
sightline impact parameter ($b$).  The measurement shown for sightline
7, which has $b= 26.9$ kpc, was made by fitting a single
Gaussian to  both absorption components reported in Table 3, i.e., it
is not a simple sum of the results from the two individual Gaussian
fits reported in  Table 3. Since the impact parameters of sightlines
9 and 10 are very similar, they are displaced from each other
in the figure for clarity. Note that the upper limits are $2\sigma$
upper limits, while the error bars are the $1\sigma$
uncertainties. The four outermost data points are suggestive of an
overall decrease of $W_0^{\lambda2796}$ with increasing impact
parameter. Quasar absorption line studies of large samples of
absorber-galaxy pairs have shown this to be true as well (Chen et
al. 2010; Rao et al. 2011).

For comparison, Figure 14 includes results from the Rao et al. (2011)
sample of absorbing galaxies which have been identified for
\MgII-selected DLAs, subDLAs, and Lyman limit systems (LLSs). The mean
redshift of the Rao et al. sample is $z\sim 0.5$, with redshifts in
the range $0.1\la z \la 1.0$. The identified absorbing
galaxies in the Rao et al. sample also have a range of luminosities,
mostly $0.1 \la L \la 1.0 L^*$, but there is not a
significant correlation between luminosity and impact parameter. Rao
et al. found only a marginal  (1.8$\sigma$) correlation between
$W_0^{\lambda2796}$ and $b$.  The solid black circles in Figure 14 are
DLAs and the open circles are subDLAs and LLSs. The data from this
current M31 study are in red. Sightlines 5 through 10
have averaged integrated 21 cm  emission \HI\ column densities in the
subDLA regime, with the exception of the Galactic and M31 blended component 
along sightline 10. (See \S 3.)  \HI\ 21 cm emission maps are not available as far out
as the four outermost sightlines, but since the $N_{HI} = 1.9\times 10^{18}$
cm$^{-2}$ edge of the \HI\ disk
of M31 is at $b\approx 33$ kpc (Figure 1),  these
sightlines are not expected to have averaged integrated \HI\ column
densities in the DLA or subDLA regime.

\begin{figure*}
\vspace{-2.0in}
\includegraphics[width={1.\textwidth}]{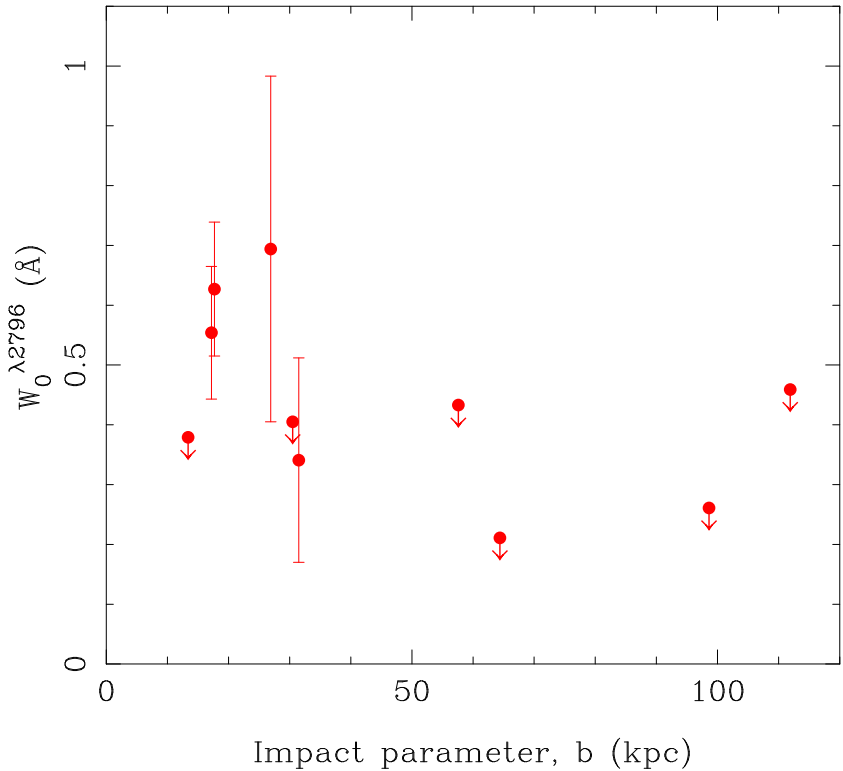}
\vspace{-3.0in}
\caption{\MgII\ $\lambda2796$ rest equivalent width,
$W_0^{\lambda2796}$,  vs. impact parameter, $b$, for M31 measurements
from Table 3.  Detections have $1\sigma$ error  bars and  arrows
indicate $2\sigma$ upper limits for the non-detections. For sightline 7,
which is the data point at  26.9 kpc, a single Gaussian fit solution
to the HVC and M31 components is  shown. It has
$W_0^{\lambda2796}=0.694 \pm 0.289$ \AA. The two points at $b\sim 17.5$
kpc have been displaced for clarity. The blended Galactic and M31
absorption along sightline 10 with $W_0^{\lambda2796}=1.096 \pm 0.122$ \AA\
has been excluded.}
\end{figure*}

\begin{figure*}
\vspace{-2.0in}
\includegraphics[width={1.\textwidth}]{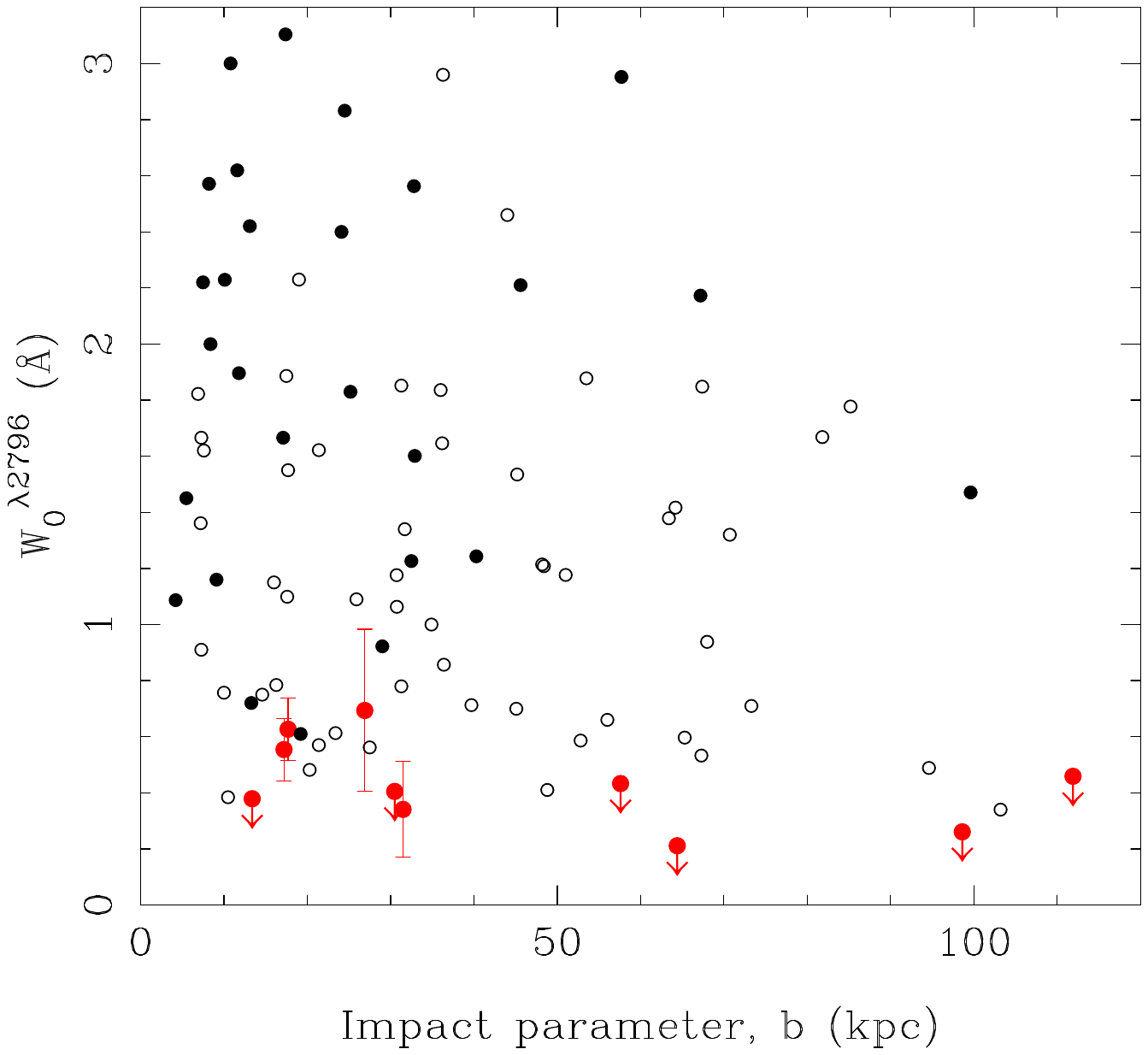}
\vspace{-3.0in}
\caption{Same as Figure 13, but data points from  Rao et al. (2011)
have been added. These represent identified galaxy  impact parameters
for \MgII\ systems with \HI\ column density measurements at $z\sim
0.5$.  Solid black circles are the DLAs as measured in UV spectra 
(Rao, Turnshek, \& Nestor 2006) and open black circles are
subDLAs and LLSs.}
\end{figure*}

Thus, as noted in \S4.1 and \S4.2,  it is clear that the sightlines
passing near M31, or  through its gaseous disk seen in 21 cm emission,
do not give rise to the moderate-to-strong \MgII\ absorption lines
which are often identified in  moderate- to high-redshift quasar
absorption-line surveys.  For
comparison, all of the Galactic detections reported in Table 3 have
$W_0^{\lambda2796}> 0.5$ \AA, and 4 of the Galactic  sightlines have
$W_0^{\lambda2796}\sim 1$ \AA\ (sightline 10 is a blend of Galactic and M31 gas).  
In the HST Key Project sample of
Galactic sightlines (Savage et al. 2000) the median value is
$W_0^{\lambda2796}=1.17$ \AA,  and the strongest line has
$W_0^{\lambda2796}=2.2$ \AA.

Of course, our sightlines through M31 are biased sightlines in the context of
traditional absorption line surveys, in that the galaxy was pre-selected in
order to study the properties of its low-ion and high-ion gas.  Therefore, 
for M31 the probability of occurrence of \MgII\
absorption as a function of $W_0^{\lambda2796}$ is not properly
estimated from the observed incidence of \MgII\ absorption in unbiased
quasar absorption-line surveys.  Instead, however, this experiment
does show that a gas-rich, $\sim 2L^*$, spiral galaxy like M31 need not
give rise to moderate-to-strong \MgII\ absorption along sightlines
which pass through its \HI\ 21 cm emission disk, or even through a
putative extended gaseous halo.

\subsection{Comparison of 21 cm Emission and Absorption-Line Velocities}

The range of velocities that exhibit 21 cm emission for sightlines
5 through 10 are shown as cyan and orange vertical bars as a function of
impact parameter in Figure 15. Cyan bars correspond to 21 cm emission velocities from M31 gas
and orange bars represent HVC velocities.  Also plotted are the  velocities of
the low-ion (red stars) and high-ion (blue triangles) absorption
lines from Table 4. The Galactic and M31 blended low-ion absorption line
along sightline 10 is shown as the encircled star. For the two inner disk sightlines  (9 and
10 at $\approx17.5$ kpc),  it appears that the velocity of the
high-ion absorption is better correlated with the 21 cm emission
velocity range.  Along sightline 8, the low and high-ion
absorption lie at an outer velocity edge of where 21 cm emission is
detected. As noted in \S3, this velocity corresponds to the peak in 21
cm emission along this sightline.  For sightlines 7 (at
$b=26.9$ kpc) and 5 (at $b=31.5$ kpc), the low-ion gas again coincides
with the peak of 21 cm emission which is near near the edge of the 21
cm profile (see Figures 7 and 9).   Thus, in nearly all cases, the low
ions occur near the edge of the 21 cm profiles (two are near the low
velocity edge and three are near the high velocity edge), and for
sightlines 5, 7, and 8, are coincident with the peak in 21 cm emission.

\begin{table*}
\begin{center}
\caption{Heliocentric velocity offsets of low- and high-ion absorption lines\tablenotemark{a}}
\begin{tabular}{rcccc}
\tableline\tableline \multicolumn{1}{c}{Quasar} & \multicolumn{2}{c}{M31} & \multicolumn{2}{c}{Milky Way}\\
 & Low ion & High ion & Low ion & High ion \\
 & (km s$^{-1}$) & (km s$^{-1}$) & (km s$^{-1}$) & (km s$^{-1}$) \\
\noalign
{\hrule}
1. 0018+3412 & \nodata & \nodata & $-18$ & $-61$ \\
2. 0024+3439 & \nodata & \nodata & $-6$ & \nodata \\
3. 0030+3700 & \nodata & \nodata & $-21$ & $-53$ \\
4. 0031+3727 & \nodata &  \nodata & $-33$ & \nodata\\
5. 0032+3946 & $-453$ & \nodata & $-42$ & \nodata\\
6. 0037+3908 & $-508$ & \nodata & $-42$ & \nodata\\
7. 0040+3915 & $-513$,$-389$\tablenotemark{b} &\nodata & $-38$ & \nodata\\
8. 0043+4016 & $-336$ & $-340$ & $6$ & $42$ \\
9. 0043+4234 & $-234$ & $-191$ & $-73$ & $-1$ \\
10. 0046+4220 & $-195, -40$\tablenotemark{c} & $-152, -35$\tablenotemark{c} & $-40$ & $-35$ \\
\tableline
\end{tabular}
\tablenotetext{a}{The velocity centroid of the Milky Way absorption system is determined from the 
\MnII\ $\lambda2576$ line, if detected, or from the \MgII\ $\lambda2796$ line if no \MnII\ is present,
or from the \CII\ $\lambda 1334$ if neither is present in 
the spectrum. The velocity centroid of the  \CIV\ $\lambda1548$ line was determined independent
of the low-ion velocity, and was used to constrain the positions of the high-ionization lines. 
The uncertainties in the low- and high-ion velocities are 6 km s$^{-1}$ and 16 km s$^{-1}$, respectively.}
\tablenotetext{b}{The two measurements are M31 HVC and halo components, respectively. See Figure 9.}
\tablenotetext{c}{The two measurements are M31 halo and disk components, respectively. The disk component
is blended with the Milky Way line.}
\end{center}
\end{table*}

\begin{figure*}
\vspace{-2.0in}
\includegraphics[width={1.0\textwidth}]{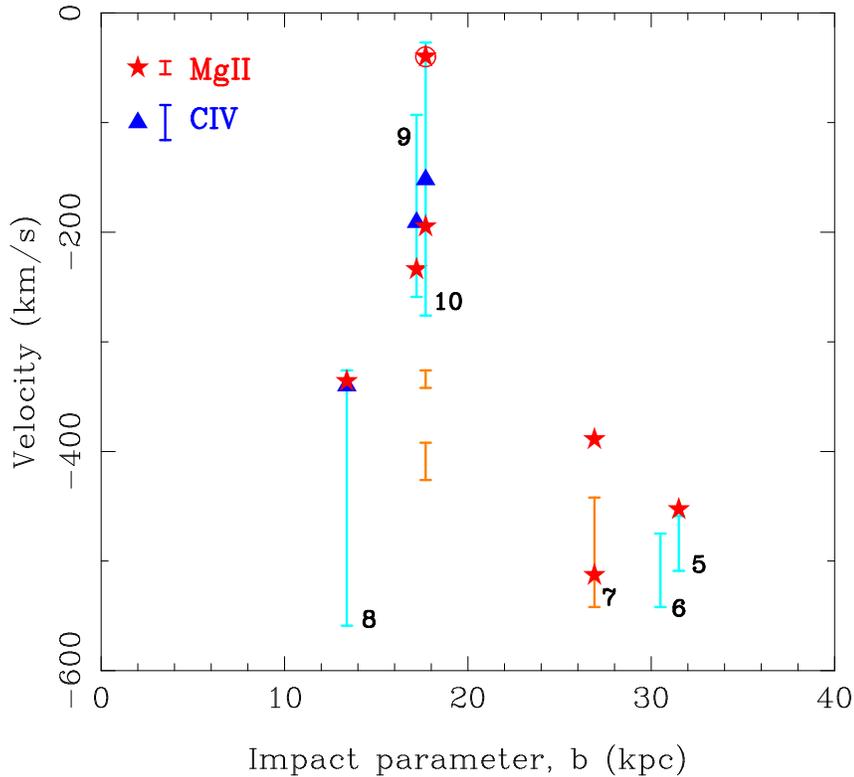}
\vspace{-3.in}
\caption{Velocities of detected lines in M31 as a function of impact
parameter.  Cyan and orange vertical bars represent velocity ranges of 21 cm
emission from M31's disk and HVCs, respectively. (See the bottom panels of Figures 7-12
for an indication of the velocity regimes which contain the most gas.)  Red stars
are low-ion (\MnII, \MgII, or \CII) line centroids, and blue triangles
are high-ion (\CIV) line centroids from Table 4.  The uncertainty in
the velocity measurement is shown as the vertical bar in the upper
left corner.  Sightlines 9 and 10 are displaced for clarity.  Three
distinct velocity ranges are apparent towards sightline 10; the wide
component arises in M31, and is partly blended with Milky Way gas. 
The encircled star at $-40$ km s$^{-1}$ is the blended Galactic and M31 disk
absorption-line velocity centroid, and the red star at $-195$ km s$^{-1}$ is from M31's halo.
 (See description of sightline 10 in \S 3.) 
The two narrower orange components originate  in M31 HVCs. No metal lines are detected at 
these velocities. The two red stars along sightline 7 are the
two components of the \MgII\ line shown in Figure 9. 21 cm emission is
detected only from the HVC along this sightline but not at the
velocity of the \MgII\ component at $-389$ km s$^{-1}$. We therefore
surmise that  this gas resides in the halo and not in the disk of M31.
We caution that the velocities plotted in this figure are not a measurement of M31's rotation curve
since, except for sightline 10, the inner sightlines, i.e., 5-9, 
do not lie along the major axis of M31. See Figure 1.}
\end{figure*}

Given the resolution of the NUV and FUV data ($\sim 87$ km s$^{-1}$ at
$\sim 2800$ \AA\ and $\sim 106$ km s$^{-1}$ at $\sim 1550$ \AA), one
might question if these differences  are significant. However, it is
well-known that  in data with sufficient signal-to-noise, a  Gaussian
fit to an absorption line can be used to determine the centroid
location of the line to an accuracy much better than the line's FWHM.
In order to determine how accurately absorption-line locations can be
determined, we ran 10,000 realizations of lines with equivalent widths
drawn from the  data. Figure 16 shows the distributions of equivalent
widths.  Noise was added to the Gaussian profiles generated with these
equivalent widths so that the resulting signal-to-noise ratios matched
the data. Line centroids  were then estimated by refitting Gaussian
profiles to the noised-up absorption lines. The resulting
distributions of centroid velocities relative to the input  values are
shown in Figure 17. For both the original as well as the simulated
data, the spectra were rebinned to two pixels per resolution element
before measurements were made. The signal-to-noise ratios of NUV
spectra were, in general, higher than in FUV spectra. Thus, the
accuracy with which the line centroids can be measured is higher  for
the \MgII\ lines. Specifically, the centroid standard deviation of the
\MgII\ distribution is $\sim6$ km s$^{-1}$ compared to $\sim16$ km
s$^{-1}$  for \CIV. These uncertainties indicate that the separations
in velocities of the low and high ions are significant towards
sightlines 9 and 10 at approximately the $2\sigma$ level.

\begin{figure*}
\includegraphics[width=3.0in]{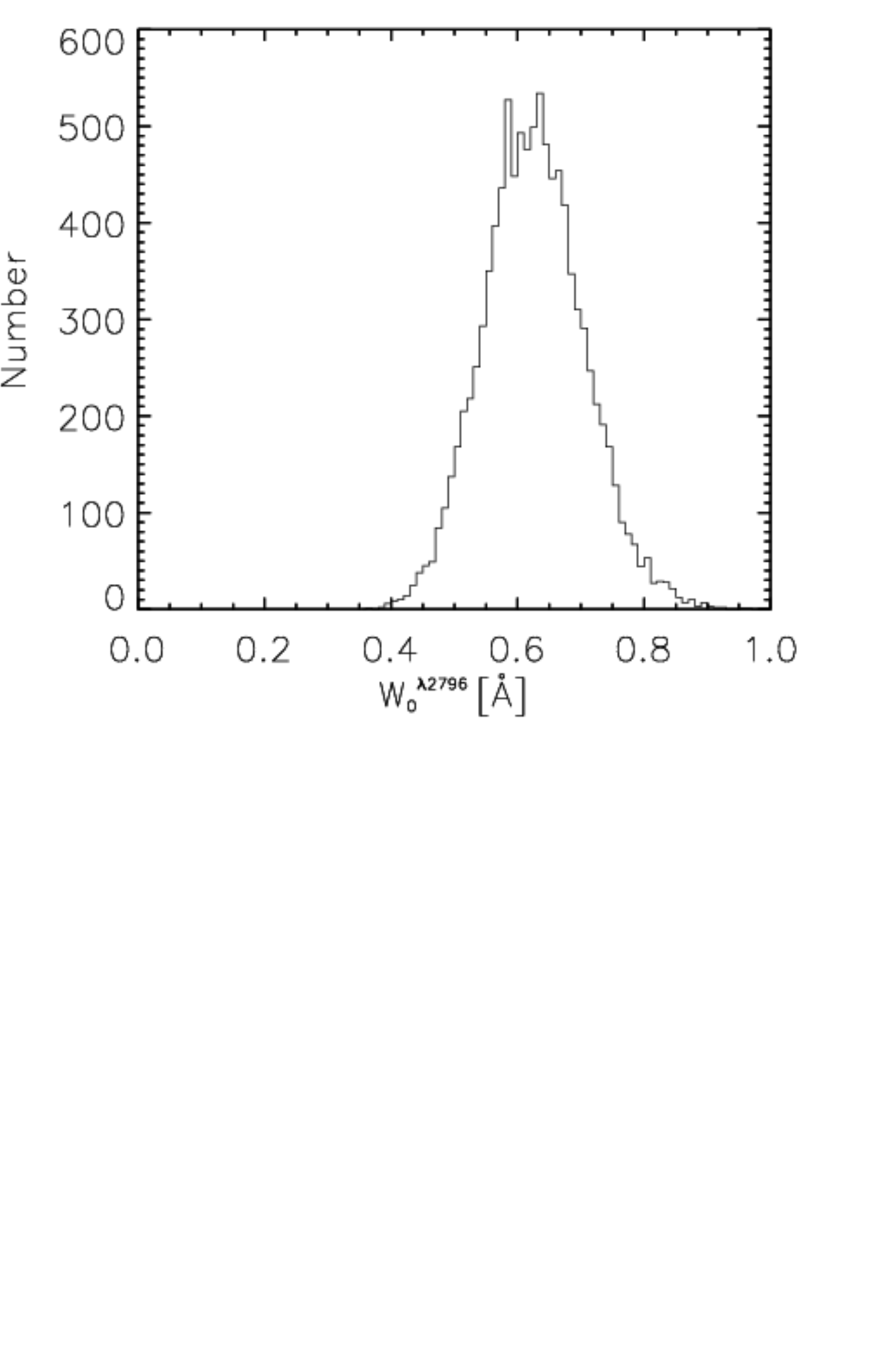}
\includegraphics[width=3.0in]{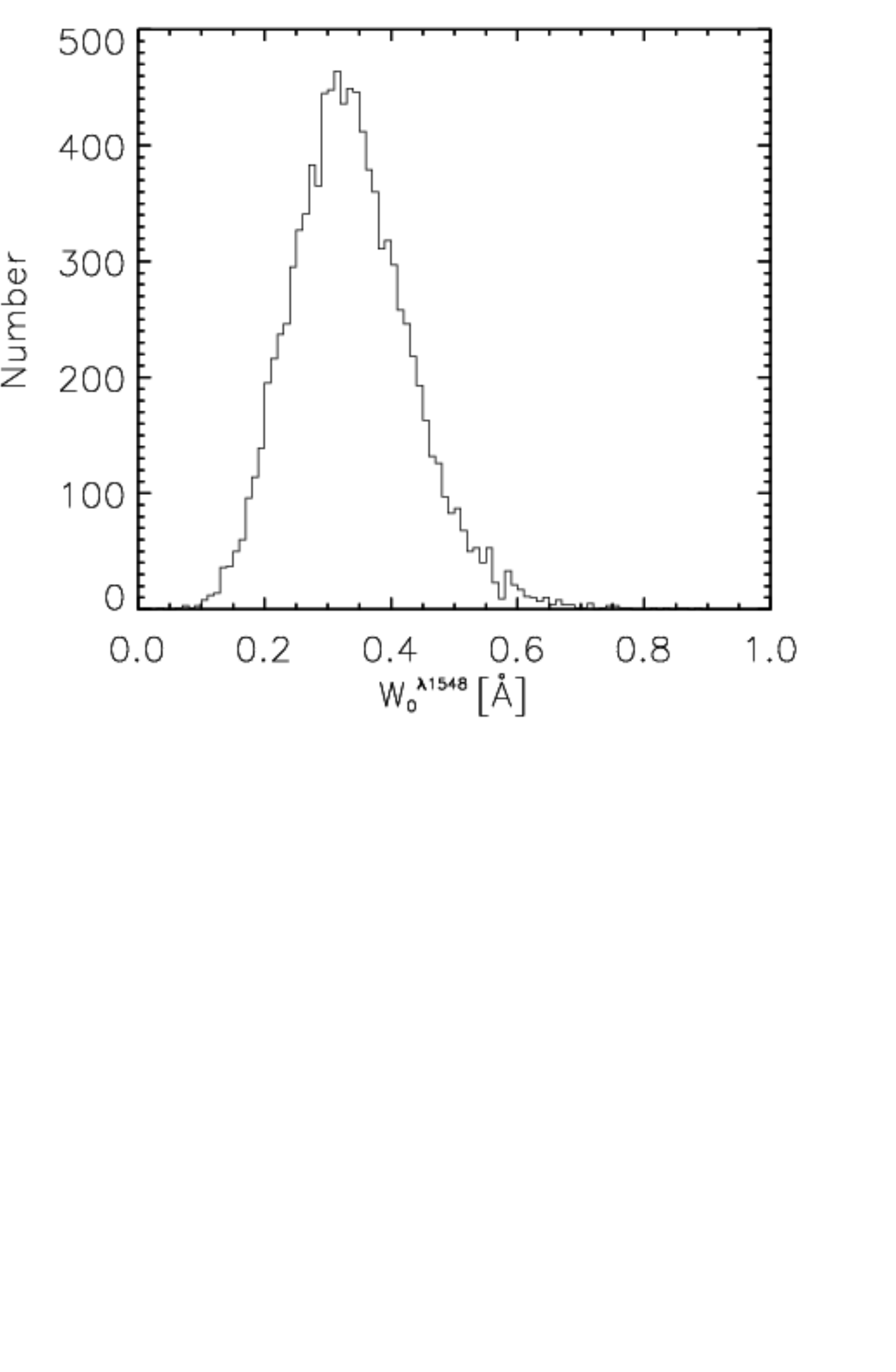}
\vspace{-2.0in}
\caption{Distribution of rest equivalent widths, \MgII\
$W_0^{\lambda2796}$ (left) and  \CIV\  $W_0^{\lambda1548}$ (right),
from 10,000 realizations of the data. }
\end{figure*}

\begin{figure*}
\includegraphics[width=3.0in]{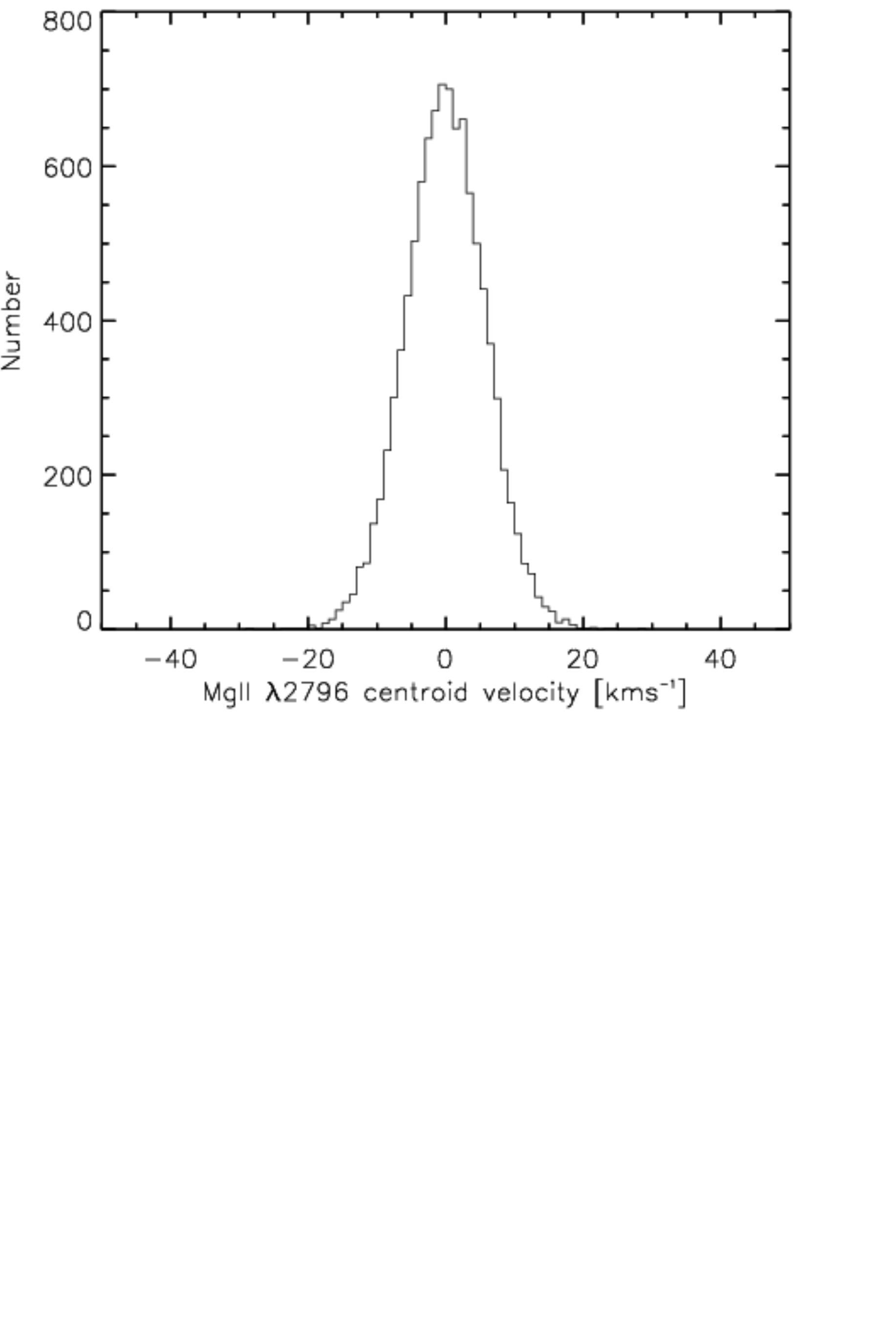}
\includegraphics[width=3.0in]{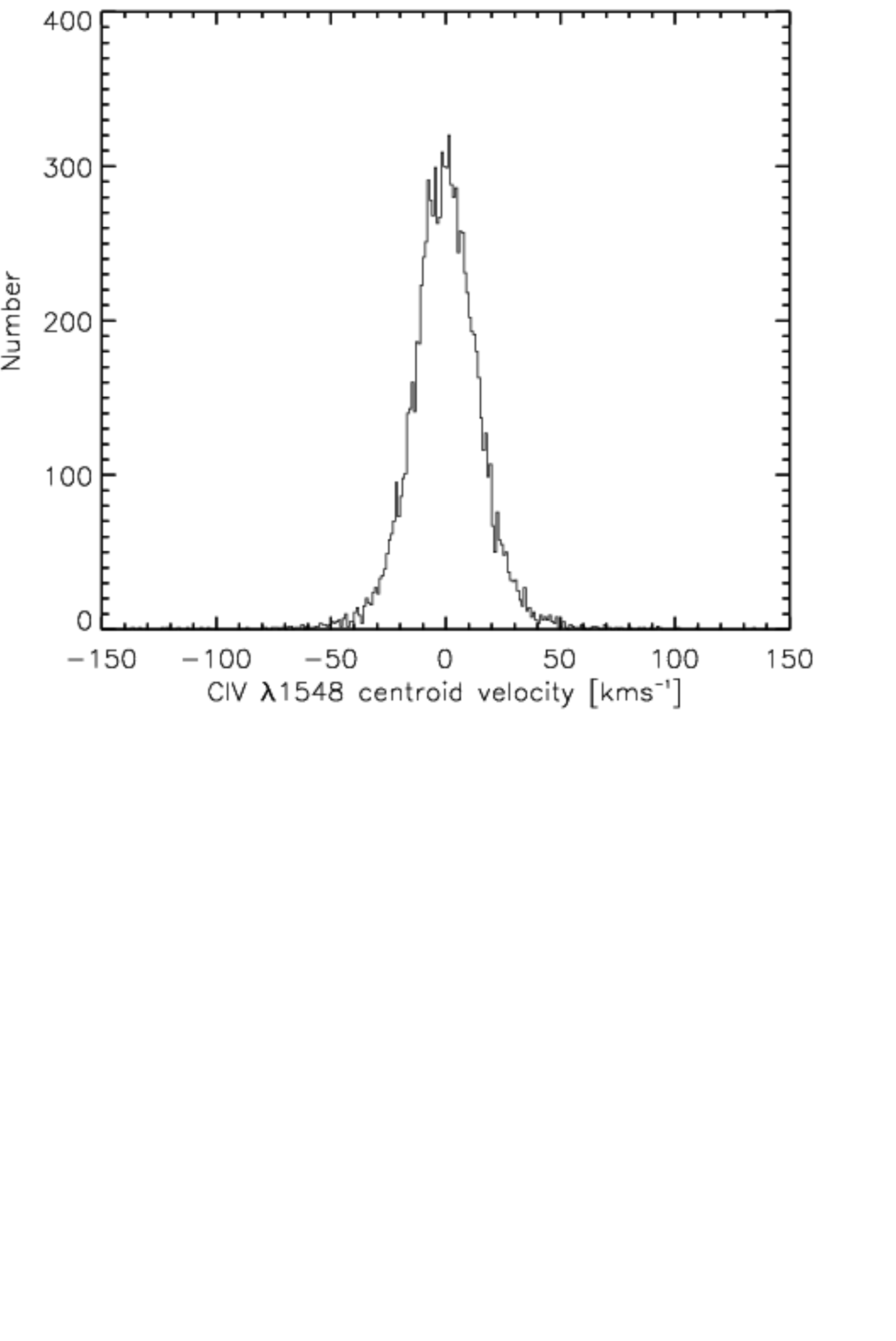}
\vspace{-2.0in}
\caption{Distribution of line centroid velocity offsets measured from
10,000 realizations of the data. Gaussian profiles with rest
equivalent widths sampled from measured values were generated, to
which noise was added to  match the signal-to-noise ratio of the
data. Centroid velocities of these simulated lines were measured, and
the offsets from input values are shown here. We report the standard
deviation of this distribution as the uncertainty in  the centroid
velocity measurement, i.e., the \MgII\ $\lambda2796$ and \CIV\
$\lambda1548$ line centroids can be measured with an accuracy of 6
km s$^{-1}$ and 16 km s$^{-1}$, respectively.  }
\end{figure*}

The 21 cm emission studies of M31 (e.g., \S2.1) show that for this
nearly edge-on galaxy, the sightline velocities of gas giving rise to
21 cm emission can span a large range (e.g., see the lower panels in
Figures 7 - 12).  Corbelli et al. (2010) fitted a tilted ring model to
M31's \HI\ 21 cm emission data from 8 to 37 kpc to study the details
of its rotation, finding that  M31's disk warps beyond galactocentric
distances of $\sim 25$ kpc and that it becomes more  inclined with
respect to our sightline.   As we have shown above, the \MgII\
absorption regions are almost always at the peak of the 21 cm emission 
profile, which occurs near the edge of the 21 cm
emission velocity range. Thus, when detected, the low-ion gas appears to
trace the 21 cm gas.  
Interestingly, neither low- nor high-ion absorption lines
are  detected at the 21 cm velocity locations of the HVCs along
sightline 10 ($b=17.5$ kpc). Low-ion absorption is also not
detected at the 21 cm disk velocity location towards sightline 6
($b=30.5$ kpc). However, the observed low-ion absorption along
sightline 7 originates in the HVC  detected in 21 cm emission,
but at the velocity location of the other  absorption component,
there is no detected 21 cm emission.  This component, at $-389$ km
s$^{-1}$, is likely to be M31 halo gas.  Thus, it appears that the
sightlines through M31 are passing through very different physical and
kinematic conditions within its ISM.

\section{Conclusions}

A conventional study relating quasar absorption-lines to the galaxies
that cause them begins with the detection of an intervening
absorption-line system in a spectrum followed by imaging work to
identify the galaxy.  The experiment with M31 described here is a
quasar absorption-line survey conducted in reverse.  We probed ten
sightlines with vastly different impact parameters through a single
spiral galaxy with a luminosity of  $\sim 2L^*$. As summarized  in \S4.1,
we detected some type of absorption from M31 gas in five of the six
inner sightlines ($13 < b <32$ kpc), but no absorption  in any of the
four outer sightlines ($57 < b < 112$ kpc). We also reported the first
detection of metals in a M31 HVC. 

In \S4.3 we compared our M31 results to the findings in the
conventional Rao et al. (2011) survey. Rao et al. found only a
marginal anticorrelation between $W_0^{\lambda2796}$ and $b$, and
indeed we find the same qualitative trend in M31, but the values of
$W_0^{\lambda2796}$ are far smaller in M31 (Figure 14).  And while Rao
et al. found that there were fewer systems with moderate-to-strong
$W_0^{\lambda2796}$ at large-$b$ ($b>50$ kpc), we found none arising
in M31. In \S4.4 we compared the velocity locations of low-ion and
high-ion gas in M31 to that of M31's  21 cm emission and found that
the high-ion gas is better aligned with the velocities of observed 21
cm emission along two of three sightlines where it is detected.  The
velocity of the low-ion gas is correlated with the peak of 
21 cm emission and is often near the edge of the 21 cm
emission velocity range. In one case \MgII\ is detected at a velocity
location that  shows no 21 cm emission.

Broadly, our results indicate that:

\begin{enumerate}

\item Despite the fact that M31 is a gas-rich, $\sim 2L^*$ spiral galaxy,
it produces relatively weak \MgII\ and \CIV\ absorption lines in
comparison to those found in moderate-to-high redshift quasar
absorption-line surveys. For \MgII, this may indicate that M31 is
typical of a class of luminous galaxies that don't possess gaseous
cross sections capable of giving rise to  moderate-strength quasar
absorption lines even at impact parameters $b \la 32$ kpc. This
finding might also be related to the observed relative decrease in the
incidence of stronger \MgII\ systems with decreasing redshift.

\item M31 appears not to possess an extensive large gaseous cross
section at impact  parameters $b>57$ kpc that is capable of giving
rise to  moderate-strength quasar absorption lines (e.g., with
$W_0^{\lambda1548} > 0.2$ \AA\ or $W_0^{\lambda2796} > 0.3$ \AA), at
least not along the direction of its major axis.

\item For the relatively weak absorption that we did detect at $b
\la 32$ kpc, we found the low-ion gas to be associated with the
peak in the 21 cm emission profile, near one edge  of the 21 cm
emission velocity range. Two of three sightlines showed high-ion  gas to be
centrally located within the 21 cm emission profile, with the  third
being coincident with an edge. It is also likely that we have detected
low-ion halo gas through two of the sightlines. 

\end{enumerate}

Future UV spectroscopy of quasars behind M31 can build on these
findings by: (1) acquiring higher signal-to-noise data to probe down
to weaker rest equivalent width values, (2) acquiring higher
resolution data to better study the velocity locations of the gas
relative to 21 cm emission velocities, and/or (3) probing a larger
number of sightlines including ones in M31's extended halo region.
 
It would be interesting if $N_{HI}$ results derived from M31's 21 cm
emission data could be compared with $N_{HI}$ determinations from
Lyman series absorption seen in the UV spectra of background quasars.
One could then get an \HI\ column density measurement averaged over
less than a milli-parsec region in M31, in comparison to the $\sim$ 50
pc linear spatial scale offered by the radio observations. This would
provide information on the homogeneity and size scale of \HI\
absorbing regions in M31.

\section*{Acknowledgments}

We are grateful for the referee's encouraging remarks and 
comments which improved and clarified the paper. 
SMR, DAT, RW, DT, and DVB acknowledge support from HST grant GO-11658.
GMS acknowledges support from a Zaccheus Daniel Fellowship and a 
Dietrich School of Arts and Sciences Graduate Fellowship from the 
University of Pittsburgh. We are grateful for the help and support 
provided by the NOAO staff. We thank Mich\'ele Belfort-Mihalyi for 
assisting on the NOAO observing run.

{}

\label{lastpage}
\end{document}